\documentclass[12pt]{article}
\pdfoutput=1
\usepackage{jheppub}
\newcommand{\be}{\begin{equation}}
\newcommand{\ee}{\end{equation}}
\newcommand{\lan}{\langle}
\newcommand{\ran}{\rangle}
\newcommand{\Ss}{\mathcal{S}}
\newcommand{\Nn}{\mathcal{N}}

\begin{document}


\title{Inflation after False Vacuum Decay: \\ Observational Prospects after Planck} 

\author[a,b]{Raphael Bousso,}
\author[c]{Daniel Harlow,}
\author[d,e,f]{and Leonardo Senatore}
\affiliation[a]{Center for Theoretical Physics and Department of Physics,\\
 University of California, Berkeley, CA 94720, U.S.A.}
\affiliation[b]{Lawrence Berkeley National Laboratory, Berkeley, CA 94720,
  U.S.A.}  
\affiliation[c]{Princeton Center for Theoretical Science, Princeton University, Princeton NJ 08540 USA} 
\affiliation[d]{Stanford Institute for Theoretical Physics,
 Stanford University, Stanford, CA 94306} 
\affiliation[e]{Kavli Institute for Particle Astrophysics and Cosmology, \\
Stanford University and SLAC, Menlo Park, CA 94025} 
\affiliation[f]{CERN, Theory Division, 1211 Geneva 23, Switzerland}

\emailAdd{bousso@lbl.gov}
\emailAdd{dharlow@princeton.edu}
\emailAdd{senatore@stanford.edu} 

\abstract{We assess two potential signals of the formation of our universe by the decay of a false vacuum. Negative spatial curvature is one possibility, but the window for its detection is now small.  However, another possible signal is a suppression of the CMB power spectrum at large angles. This arises from the steepening of the effective potential as it interpolates between a flat inflationary plateau and the high barrier separating us from our parent vacuum.  We demonstrate that these two effects can be parametrically separated in angular scale.

Observationally, the steepening effect appears to be excluded at large $\ell$; but it remains consistent with the slight lack of power below $\ell\approx 30$ found by the WMAP and \textsl{Planck} collaborations.  We give two simple models which improve the fit to the \textsl{Planck} data; one with observable curvature and one without.  Despite cosmic variance, we argue that future CMB polarization and most importantly large-scale structure observations should be able to corroborate the \textsl{Planck} anomaly if it is real.

If we further assume the specific theoretical setting of a landscape of metastable vacua, as suggested by string theory, we can estimate the probability of seeing a low-$\ell$ suppression in the CMB. There are significant theoretical uncertainties in such calculations, but we argue the probability for a detectable suppression may be as large as ${O}(1)$, and in general is significantly larger than the probability of seeing curvature. 
}
\maketitle

\section{Introduction and Summary} 
\begin{figure}
\begin{center} 
\includegraphics[height=6.5cm]{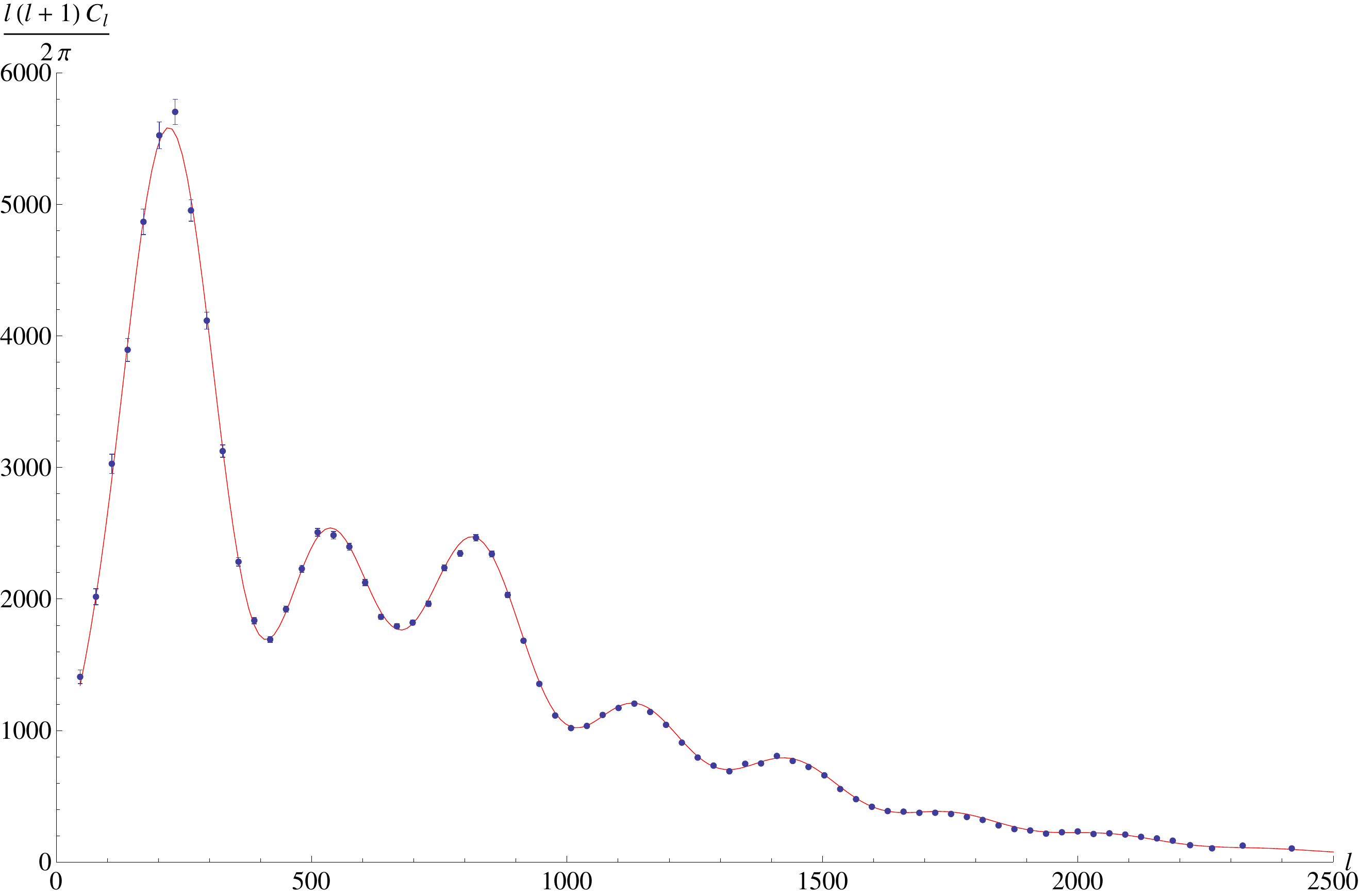} \end{center} \caption{\textsl{Planck} data for $\frac{\ell(\ell+1)}{2\pi}C_\ell$, in $(\mu K)^2$, for $\ell\geq50$.  The red theory curve is $\Lambda$CDM with the \textsl{Planck} best-fit parameters, as computed using CLASS \cite{Lesgourgues:2011re,Blas:2011rf,Lesgourgues:2011rg}.}\label{datafig1} 
\end{figure}

Inflation \cite{Guth:1980zm,Linde:1981mu,Albrecht:1982wi,Starobinsky:1980te} is a powerful framework for understanding the early universe, in particular the spectrum of density perturbations that led to structure formation and to anisotropies in the cosmic microwave background (CMB) \cite{Mukhanov:1981xt,Hawking:1982cz,Starobinsky:1982ee,Guth:1982ec,Bardeen:1983qw}.  Measurement of these anisotropies \cite{Smoot:1992td,Bennett:1996ce,Miller:1999qz,Spergel:2003cb,Hinshaw:2012aka} out to $\ell\approx 1000$ were crucial in establishing the standard $\Lambda$CDM model of cosmology.  Very strong additional evidence for this model has recently been provided by the \textsl{Planck} collaboration \cite{Ade:2013xsa,
Planck:2013kta,Ade:2013lta,Ade:2013uln}, which reported measurements of temperature anisotropies out to $\ell\approx 2500$.  Their results for the power spectrum above $\ell=50$ are shown in Fig.~\ref{datafig1}.

There are some basic facts in cosmology which are not explained by the $\Lambda$CDM model.  Chief among these is the smallness of the observed value of the cosmological constant.  Moreover, it is not obvious that slow-roll inflation is generic. One could argue that some tuning of both the fundamental theory and the initial conditions is required: 
\begin{itemize}
\item {\em Theory Parameters:} For the observed structure to be seeded, there must be a hierarchy between the fundamental scale and the scale of inflation, and also sufficient flatness over a large enough field range. 
\item {\em Initial Conditions:} For inflation to begin, vacuum energy must dominate in a region of order the associated Hubble scale. If initial conditions can be tuned to give rise to such a patch, why not solve the horizon, homogeneity, and flatness problems directly by tuning?  Moreover why should the field start off high enough on the potential to inflate but not so high as to overshoot the inflationary region?
\end{itemize} 
 It is difficult to quantify such concerns except within a specific and well understood theoretical setting.

A setting in which one could aspire to do this is the string landscape of metastable vacua \cite{Bousso:2000xa,Kachru:2003aw,Susskind:2003kw,DenDou04b}, populated by eternal inflation \cite{Steinhardt:1982kg,Vilenkin:1983xq,Linde:1986fd}. In this setting one can in principle derive statistical predictions for the parameters of $\Lambda$CDM and inflation. This determines which regions of parameter space are typical and which are not.  

In practice, our understanding of the string landscape is too primitive to derive prior distributions from the top down.  For a few parameters the prior can be derived with some confidence. In particular, the cosmological constant should have a flat distribution near $\Lambda=0$ if supersymmetry is broken~\cite{Weinberg:1987dv}.  Along with the assumption that observers require galaxies, this can explain its smallness~\cite{Sak84,Linde:1984ir,Banks:1984cw,Weinberg:1987dv,Bousso:2000xa}. (With the causal patch measure, specific anthropic assumptions can be eliminated and, in addition, the coincidence that vacuum energy is comparable to the present matter density can be explained~\cite{BouHar07}.)  Other parameters can also be treated using simple phenomenological assumptions about their distributions~\cite{Vilenkin:1996ar,Tegmark:1997in,Freivogel:2005vv,Freivogel:2008qc,Bousso:2013rda}.

In the string landscape, our universe would emerge through the decay of a metastable de~Sitter vacuum, as a Coleman-De Luccia bubble \cite{Coleman:1980aw}. This setting goes a long way towards solving the initial conditions problem for inflation. The $SO(3,1)$ invariance of the bubble ensures a homogeneous, isotropic FRW universe inside. Because neighboring vacua typically have very different cosmological constant, the potential energy after decay can be large enough for slow-roll inflation and reheating to follow.  And the spatial curvature of the universe inside is always negative, which creates Hubble friction that prevents the field from overshooting the inflationary plateau~\cite{Freivogel:2005vv}. 

\begin{figure}
\begin{center}
\includegraphics[height=6cm]{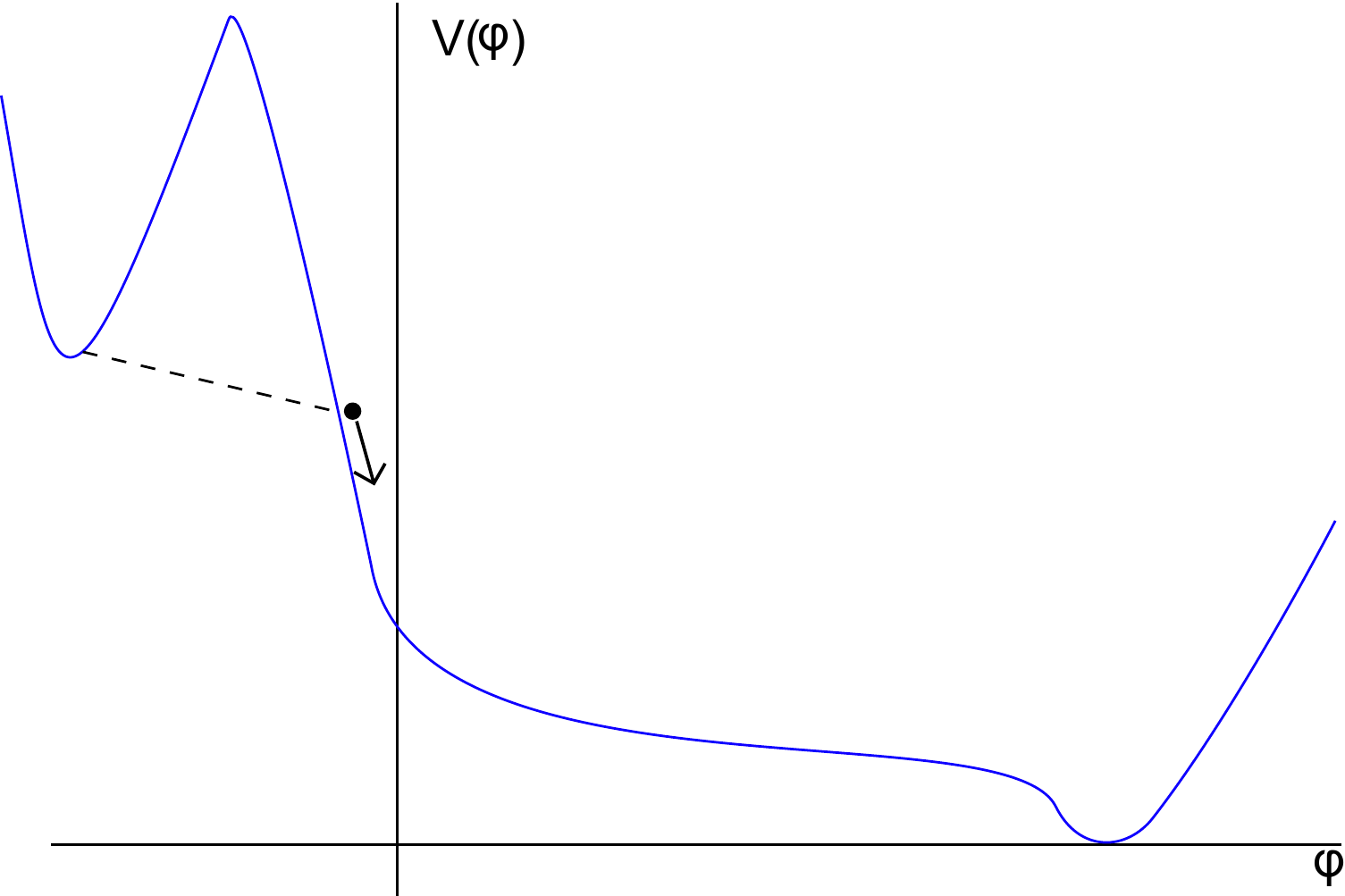}
\end{center}
\caption{The decay of our parent vacuum followed by slow-roll inflation.  The dashed line indicates a tunneling event from the high energy false vacuum, after which the field classically rolls down the hill.  There is an initial period of curvature domination, where $a\approx t$, during which the field does not travel very much.  This ends when $\frac{1}{a^2}$ becomes of order $V(\phi)$, after which inflation begins.  Eventually inflation exits and the system reheats into our vacuum, with small cosmological constant.  For later convenience we choose $\phi=0$ where modes of wavelength equal to our horizon size today are just exiting the inflationary horizon.}
\label{potentialfig}
\end{figure}

Motivated by this theoretical setting, but without committing specifically to the string landscape, we will consider the phenomenological consequences of a first-order phase transition followed by inflation.  The relevant type of potential \cite{Garriga:1998he,Freivogel:2005vv} is shown in Fig.~\ref{potentialfig}. In the first sections of the paper we examine possible signatures of this type of potential using standard inflationary techniques.
Only in the final sections will we make assumptions about the distribution of potentials in the landscape; and subject to these assumptions, we will estimate the probability that an imprint of false vacuum decay will be detected in future experiments. 

If inflation was preceded by false vacuum decay, there are two generic features which lead to potentially observable signatures:\footnote{These are not the only possible signatures associated to inflation as originating after vacuum decay. Two noteworthy others are bubble collisions~\cite{Gobbetti:2012yq} and tensor modes from the nucleation process~\cite{Yamauchi:2011qq}.  The likelihood of the former is quite difficult to estimate, and in any event it has recently been strongly constrained by the optimal analysis~\cite{Osborne:2013hea,Osborne:2013jea} (see also \cite{Feeney:2010dd,Feeney:2012hj}).  The latter only affects modes whose wavelength is of order the radius of curvature, so we can simply think of it as another manifestation of the open curvature that we discuss here.  The same can be said for the recent discussion of \cite{Liddle:2013czu}.   Other more exotic possibilities include \cite{Czech:2011aa,Salem:2012gm,vanderSchaar:2013paa}.}
\begin{itemize}
\item The negative spatial curvature inside the bubble could be seen if it is not diluted by too much inflation.  The current observational bound is $\Omega_K\lesssim10^{-2}$, but $\Omega_K$ as small as $10^{-4}$ should eventually be distinguishable from cosmic variance.  
\item During the earliest part of slow-roll inflation, the potential steepens as it interpolates from the inflationary plateau to the potential barrier.  This causes the scalar field to roll faster, which suppresses the primordial power spectrum ${\cal P}\sim \frac{H^4}{\dot{\phi}^2}$ at low angular parameter $\ell$ \cite{Contaldi:2003zv}.  Depending on the strength of this effect and on the value of $\ell$ at which it turns on, it could be detectable in the CMB and in large scale structure surveys. 
\end{itemize}
The first of these effects was discussed in detail in the wide-ranging analysis of \cite{Freivogel:2005vv}.  The possibility of a signature from the steepening of the potential was also raised, without claiming that this generically leads to power suppression relative to $\Lambda$CDM.  The idea that a steeper potential leads to a low $\ell$ power suppression was noted in \cite{Contaldi:2003zv}, outside of the context of bubble nucleation. 
 
In this paper, we will argue that in some respects the power suppression from the steepening of the potential is a more promising signature to look for than negative curvature. To do so, we need to answer the following questions:
\begin{itemize}
\item Both the power suppression and the curvature originate from the same feature, the potential barrier at the beginning of inflation.  Since curvature has been bounded at the level of $|\Omega_K|\lesssim10^{-2}$, is there still hope of observing steepening?  
\item Any suppression of the power would have to happen below $\ell \approx 50$, since the running is strongly constrained at smaller scales.  For such large angles, will cosmic variance prevent us from ever being able to confirm this feature with high confidence?
\item If false vacuum decay induces a steepening feature, what is the probability that the associated power suppression begins between $\ell=2$ and $\ell=50$, the range in which it may be detectable in the future?  Why not in the invisible region ($\ell\ll 1$), or at higher $\ell$?
\end{itemize}
We will be able to analyze the first two questions without any specific assumptions about the underlying theory. The last will require a quantitative discussion of plausible prior distributions in the string landscape, as well as a consideration of anthropic boundaries. 

Our theoretical understanding of both the prior and the catastrophic boundary is limited, so our analysis of this third question should be considered preliminary.

\paragraph{Summary of Results} Below we list our key findings:
\begin{itemize}
\item It is possible for steepening to be observable even if inflation lasted long enough to wipe out all observable traces of curvature. More generally, it will typically be the case that one should see the suppression before seeing curvature.  
\item One might worry that competing effects could make it uncertain whether potential steepening will result in a suppression or an enhancement of power~\cite{Freivogel:2005vv}. We show that suppression is the larger effect in the parametric limit of small slow roll parameters $\epsilon$ and $\eta$. We therefore conclude that suppression is the generic signal.  We also show that there should be no analogous suppression in the tensor power spectrum.  
\item The \textsl{Planck} collaboration has reported a suppression of power below $\ell\approx 30$~\cite{Planck:2013kta}, albeit with low significance ($2.5-3\sigma$).  This is such a low level of statistical significance that it does not demand a modification of $\Lambda$CDM. However, motivated by the theoretical case for a steepening feature, we exhibit two models that parametrize its inclusion. Both are found to improve the fit to the CMB power spectrum without violating the constraints on curvature.
\item The confidence in a suppression of power at low $\ell$ can be increased substantially, if such an effect is real.  Polarization measurements in the CMB will be of some help in this direction, but a more useful tool is large scale structure and possibly 21 cm radiation measurements.  We argue that such observations may eventually have the potential to increase the significance of the anomaly to as much as  $\sim 5-6\sigma$.
\item With a plausible prior for the distribution of the steepening feature among landscape vacua with slow-roll inflation, the probability for an observable effect to lie in the visible region can be as high as ${O}(1)$.\footnote{Throughout, by a probability $p\sim {O}(1)$ we mean that $p/(1-p)$ is of order unity; that is, $p$ is neither very close to $0$ nor very close to $1$.}  This is higher than the probability for seeing curvature under similar assumptions, which is ${O}(10\%)$.
\end{itemize}

\paragraph{Outline} In sections \ref{pertsec}--\ref{models}, we assume only that our universe was produced by the decay of a metastable vacuum, followed by slow-roll inflation. In Sec.~\ref{pertsec} we derive the perturbative effect of a steepening feature in the inflationary potential on the power spectrum. Sec.~\ref{sec-observation} discusses the observational status of this effect. We find it to be exluded for $\ell\gtrsim 100$ but allowed, and slightly favored, for $\ell\lesssim 30$. This anomaly has little significance for now, but we argue that future observations have the potential to eliminate or corroborate it. In Sec.~\ref{models}, we illustrate key features of the steepening effect using two models. We show that the onset of curvature can be parametrically separated from the steepening feature.  

In the final two sections we assume the existence of a large landscape of metastable vacua. Sec.~\ref{gensec} argues that a steepening feature near the beginning of inflation is statistically favored in the landscape, whereas flattening is disfavored. In particular, this disfavors scenarios such as slow-roll eternal inflation and Hawking-Moss tunneling. In Sec.~\ref{sec-probability}, we estimate the probability that the steepening feature lies in the observable region. We consider a range of plausible prior distributions for its location in the potential. We argue for a catastrophic boundary near $\ell\sim {O}(10^{4.5})$ and find that the probability for observable steepening can be as large as ${O}(1)$. After observational exclusion of a feature for $\ell\gtrsim 10^2$, we find that the probability for a feature at lower $\ell$ ranges between 10\% and 40\%, depending on our assumptions about the prior distribution.

\paragraph{Related Work} Our work builds on the seminal analysis of Freivogel, Kleban, Martinez, and Susskind~\cite{Freivogel:2005vv}.  Other work with some overlap includes Refs.~\cite{Garriga:1998he,Linde:1999wv,Contaldi:2003zv,Yamauchi:2011qq,Dudas:2012vv,Sagnotti:2013ica}.  We believe that each of the points we listed in our ``summary of results'' above is original to this paper.  Also the day before this paper was posted, \cite{Cicoli:2013oba,Pedro:2013pba} appeared on the arxiv, which give stringy realizations of a low-$\ell$ power suppression in the CMB from a steepening potential.  A preliminary version of our results was reported by DH at the ``Primordial Cosmology'' workshop at KITP on May 23rd, 2013, and are available online.  This work was although discussed further by DH at the ``Open Questions in an Open Universe'' workshop in Istanbul at Bogazici University, on August 12th 2013.

\section{Large Scale Power Suppression From Potential Steepening}
\label{pertsec}

In this section we study the effects of a steepening feature in the inflaton potential on the scalar and tensor power spectra of the CMB. Our working hypothesis is that if our universe was created through a bubble nucleation, then the inflaton potential is expected to be steeper at the beginning of slow-roll inflation, as in Fig.~\ref{potentialfig}. This can leave a detectable imprint, as we now show.\footnote{ Imprints from the steepening feature could be evaded if the Coleman-de Luccia decay deposits the field not on the steep slope but further along on the inflationary plateau.  We argue in appendix \ref{dropapp} that such potentials are not generic.}

We work in the context of slow roll inflation, where the parameters
\begin{eqnarray} 
\epsilon & \equiv & \frac{M_P^2}{2}\left(\frac{V'}{V}\right)^2~,\\
\eta & \equiv & M_P^2\frac{V''}{V}~,
\end{eqnarray}
are small compared to unity.\footnote{ Our results could also be derived in the effective field theory of inflation~\cite{Cheung:2007st,Senatore:2010wk}. This framework is largely motivated by the fact that all observations apart from the curvature are related to the perturbations themselves, and not to the background solution. However, here we wish to exhibit the connection between a perturbative signal (power suppression) and a property of the background solution (steepening). Therefore, we use a formalism that makes the background solution explicit.} We use the reduced Planck mass, $M_P^{-2}\equiv 8\pi G$. 

The scalar and tensor spectra are related to the dynamical variables ($H$, $\dot\phi$) and to the potential parameters ($V$, $V'$), evaluated at the time when modes with wavenumber $k$ exit the horizon during inflation~(see for example \cite{Weinberg:2008zzc}):
\begin{eqnarray}
\label{Ps}
k^3 \mathcal{P}_s & = & \frac{1}{2}\frac{H^4}{\dot{\phi}^2}\approx \frac{1}{6M_P^6}\frac{V^3}{V'^2}=\frac{1}{12}\frac{V}{M_P^4}\frac{1}{\epsilon}~,\\
k^3 \mathcal{P}_t & = & 4\frac{H^2}{M_P^2}\approx \frac{4}{3}\frac{V}{M_P^4}~.
\end{eqnarray}
The tensor to scalar ratio is
\be\label{req}
r\equiv \frac{\mathcal{P}_t}{\mathcal{P}_s}\approx 16\epsilon~.
\ee
In the slow-roll approximation these power spectra are close to a power-law spectrum, which for $\mathcal{P}_s$ is often parametrized as
\be\label{Ppar}
k^3\mathcal{P}_s =2\pi^2 A_s \left(\frac{k}{k_{piv}}\right)^{n_s-1}.
\ee
Here $k_{piv}$ is some reference wave number often taken to be $.05 \,{\rm Mpc}^{-1}$, and $n_s$ is called the spectral index.  The \textsl{Planck} best-fit parameters are $A_s=2.215 \times 10^{-9}$ and $n_s=.9624$~\cite{Ade:2013lta}.  For slow-roll inflation we have 
\begin{equation} 
\label{ns}
n_s-1\approx 2\eta-6\epsilon~.
\end{equation} 
To find the power spectra as functions of $k$ we need to relate $\phi$ and $k$, by solving
\be\label{phik}
\log \left( \frac{k}{k_{h}}\frac{H_{h}}{H(\phi)}\right)\equiv \Delta\mathcal{N} \approx \int_{0}^\phi\frac{d\phi}{M_P}\frac{1}{\sqrt{2\epsilon}}~,
\ee
where we used the condition that $k a_0 = a H$ at horizon crossing.\footnote{We mostly follow the conventions of \cite{Weinberg:2008zzc}, where both the scale factor $a$ and the wave vector $\vec{k}$ are dimensionful.  The dimensionless canonical conjugate to the dimensionless FRW coordinate $\vec{x}$ is $\vec{q}\equiv a_0\vec{k}$.  Unlike \cite{Weinberg:2008zzc} however, we define the Fourier transform as $f_{\vec{k}}\equiv \int d^3 x e^{-i \vec{k}\cdot \vec{x}}f(\vec{x})$.}
Here and throughout, a subscript $h$ indicates the value of a parameter at the time when scales comparable to our current horizon today were exiting the inflationary horizon.  Thus the wave number of modes which were horizon sized at that time (and therefore today) is $k_h=H_0$.  We shift $\phi$ to set $\phi_h=0$.  It will often be convenient to use the approximate relationship
\be\label{approxl}
\ell\approx k D_{ls},
\ee
which relates angular scale on the CMB to the wave number of modes which contribute strongest to the $C_{\ell}$'s at that scale.  Here $D_{ls}$ is the proper distance to the last-scattering surface today, given by
\be\label{Dls}
D_{ls}=\frac{1}{H_0}\int_{1/(1+z_{ls})}^1\frac{dx}{x^2\sqrt{\Omega_{\Lambda}+\Omega_K x^{-2}+\Omega_M x^{-3}+\Omega_R x^{-4}}}\approx \frac{3.1}{H_{0}}.
\ee

\begin{figure}
\begin{center}
\includegraphics[height=5cm]{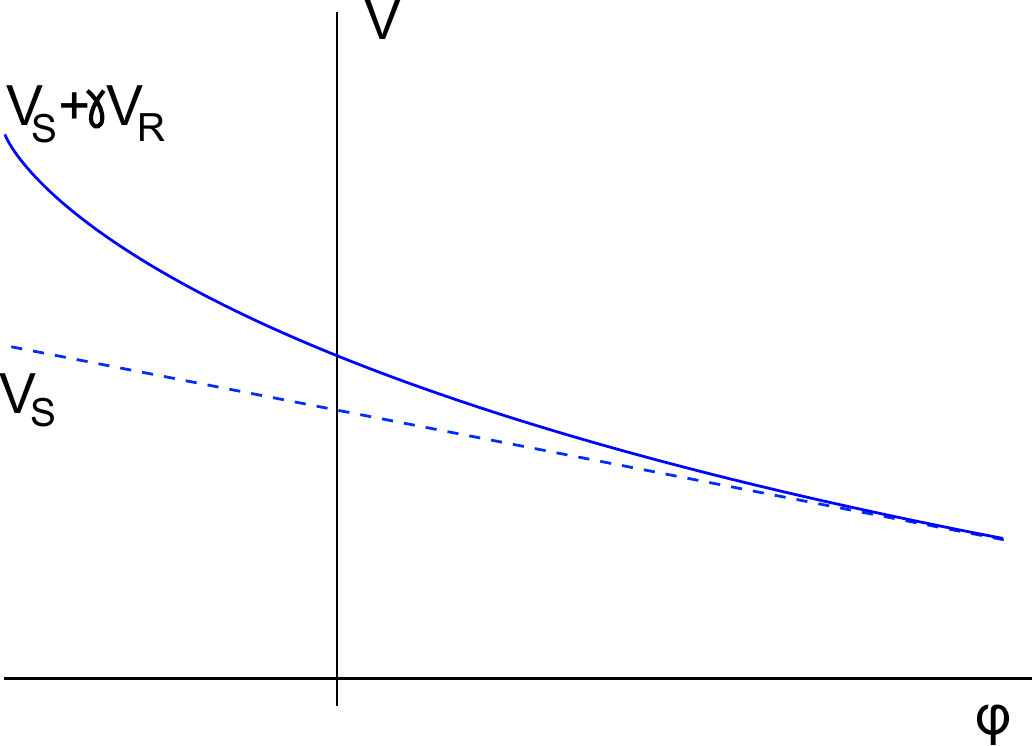}
\end{center}
\caption{Our decomposition of the potential into a flat piece $V_S$, that fits $\Lambda$CDM, and a steepening perturbation $\gamma V_R$.}
\label{steeppol}
\end{figure}

To include a steepening feature, we write the potential in the form
\begin{equation} 
V=V_S+\gamma V_R~.
\end{equation}
Here $S$ and $R$ stand for ``slow'' and ``rapid''.  $V_S$ is a slowly-varying potential whose predicted values for $n_s$, $V/\epsilon$, and $r$ are consistent with the \textsl{Planck} best-fit parameters \cite{Ade:2013lta}.  The steepening perturbation $V_R$ is a positive, monotonically decreasing function of $\phi$ that is small compared to $V_S$ at large $\phi$ but becomes order $V_S$ as $\phi$ approaches $0$ from the right.  $\gamma$ is a small parameter, which we can make unambiguous by normalizing $V_R$ such that
\be
V_S[0]=V_R[0]~,
\ee
remembering that by definition, $\phi_h=0$ corresponds to the present horizon scale.  This decomposition is illustrated in Fig.~\ref{steeppol}. We expand in $\gamma\ll 1$.

The full slow-roll parameter $\epsilon$ can be expanded as
\be
\epsilon=\epsilon_S\left[1+2\gamma\left(\frac{V_R'}{V_S'}-\frac{V_R}{V_S}\right)+O(\gamma^2)\right].
\ee
Crucially, the first $O(\gamma)$ correction will generically be larger than the second near $\phi=0$.  This is because $V_R'$ is significantly larger than $V_S'$, which is proportional to $\sqrt{\epsilon_S}$. This can be captured systematically by organizing the expansion in terms of
\begin{equation}
\tilde \gamma\equiv \frac{\gamma}{\sqrt{\epsilon_S}}~.
\end{equation}
We will therefore demand not just that $\gamma\ll1$ but also that $\tilde{\gamma}\ll1$, so that perturbative corrections to $\epsilon$ are small.  Expanding to first order in $\tilde\gamma$, we find
\begin{align}\nonumber
V&\approx V_S\ , \\
\epsilon&\approx \epsilon_S\left(1+2\gamma \frac{V_R'}{V_S'}\right)\ ,
\end{align}
and thus
\begin{align}\nonumber
\frac{\mathcal{P}_s}{\mathcal{P}_{s,S}}&\approx\left(1-2\gamma \frac{V_R'}{V_S'}\right)\ ,\\
\frac{\mathcal{P}_t}{\mathcal{P}_{t,S}}&\approx 1~.\label{spectrares}
\end{align}

This demonstrates the claimed effect: to leading order in the slow-roll parameters and $\tilde{\gamma}$, the scalar power spectrum is suppressed (never enhanced!) by the steepening perturbation. In this limit we can regard only the slope of the potential as perturbed, not the height; thus, the tensor spectrum is unaffected at this order.  This means that low-$\ell$ suppression should be seen in the TT, TE, and EE correlation functions in the CMB, which are dominated by $\mathcal{P}_s$, but not in the BB correlation function, which is controlled by $\mathcal{P}_t$.

We have expressed the power spectra as functions of $\phi$. What is actually measured is the power as a function of $k$, but this does not change our result at leading nontrivial order. To see this, we must solve \eqref{phik}, keeping track of corrections proportional to $\gamma$. Parameterizing the exact solution as

\be
\phi(k)=\phi_S(k)+ \delta\phi(k),
\ee
one has
\be\label{morecorr}
 \mathcal{P}_{s,S}(\phi(k))\approx\frac{1}{6 M_P^6}\frac{V_S[\phi_S+\gamma\,\delta\phi]^3}{V_S'[\phi_S+\gamma\,\delta\phi]^2}\approx \mathcal{P}_{s,S}(\phi_S(k))\left[1+\left(3\sqrt{2\epsilon_S}-2 \frac{\eta_S}{\sqrt{2\epsilon_S}}\right)\frac{\delta\phi}{M_P}\right].
\ee
To understand the size of these corrections, we need to find $\delta\phi$. Since $\frac{H_h}{H}=1+O(\sqrt{\epsilon})$, equation \eqref{phik} can be approximated as
\be
\log\left(\frac{k}{k_h}\right)\approx\int_{\phi_h}^\phi \frac{d\phi}{M_P}\frac{1}{\sqrt{2\epsilon_S}}\left[1-\gamma \frac{V_R'}{V_S'}\right].
\ee
Solving this equation perturbatively in $\gamma$ shows that
\be
\delta\phi(k)\approx \gamma \frac{V'_{R}}{V'_S}\sqrt{2\epsilon_S} \log(k/k_h)\ , 
\ee
so the corrections coming from this effect are smaller than the suppression term in \eqref{spectrares} by factors of either $\epsilon_S\log(k/k_h)$ or $\eta_S\log(k/k_h)$. It is important here that we are interested only in $\log(k/k_h)$ of order one, since the effect due to $V_R$ decays at higher $k$.  In practice, this means that we can just use the solution $\phi_S$ in~\eqref{phik}.

\section{Observation}
\label{sec-observation}

In this section, we consider the observational prospects of the power suppression from steepening.

\subsection{The \textsl{Planck} Anomaly}
\label{sec-exclude}

\begin{figure} 
\begin{center} \includegraphics[height=6.5cm]{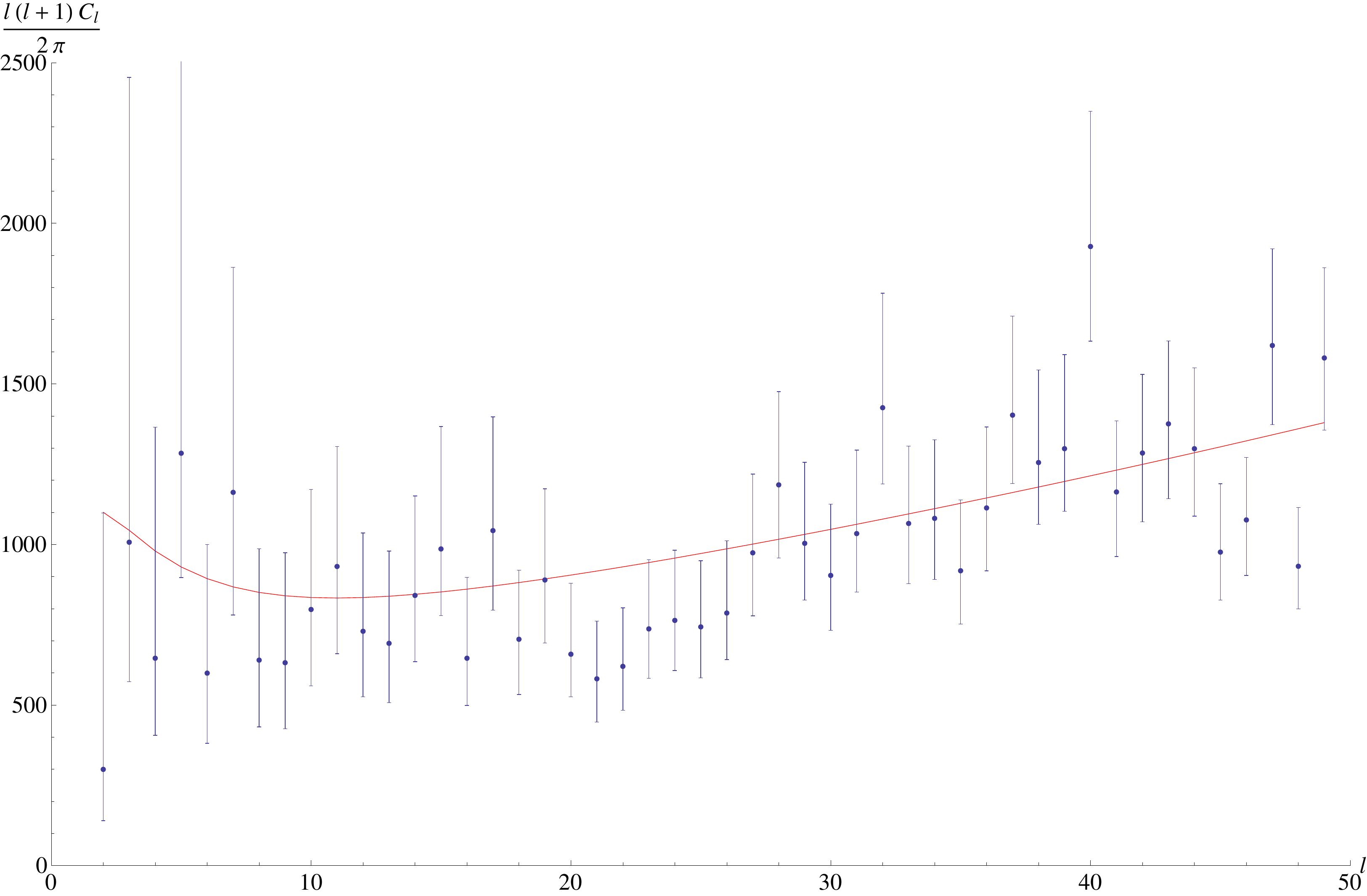} \end{center} \caption{\textsl{Planck} data for $\frac{\ell(\ell+1)}{2\pi}C_\ell$, in $(\mu K)^2$, for $\ell<50$. As in Fig.~\ref{datafig1}, the red curve is $\Lambda$CDM with the new best fit parameters~\cite{Ade:2013lta}.}\label{datafig2} 
\end{figure}

From Fig.~\ref{datafig1} it is clear that the \textsl{Planck} data is very well described by $\Lambda$CDM for $\ell\gtrsim 100$.  
The situation at low $\ell$ is less clear.
From Fig.~\ref{datafig2} it appears that for $\ell\lesssim30$ most of the $C_\ell$'s are below the best fit.


It is not clear that this suppression should be taken seriously. If the feature were localized in the middle of the power spectrum, it would be easily attributable to the look-elsewhere effect.  That it affects the lowest $\ell$'s makes it more special, but its significance is still decreased by the {\em a posteriori\/} choice of $\ell\sim 30$ for the onset of suppression.  Its significance was reported by the \textsl{Planck} collaboration~\cite{Planck:2013kta} as $2.5$\,--\,$3$\,$\sigma$ depending on the choice of estimator.\footnote{ A similar suppression was already seen in WMAP, at even lower significance.  With current data analysis by the {\textsl{Planck}} team~\cite{Planck:2013kta}, the measurement of the low $C_\ell$'s by \textsl{Planck} is not appreciably more precise than that of WMAP; both are essentially limited by cosmic variance. However, the inclusion of the new measurements at high-$\ell$ by \textsl{Planck} gives $\Lambda $CDM less flexibility to fit the low $C_\ell$'s.  In other words, the data points have not gone down, but the best fit curve has gone up.  This has the practical effect of increasing the significance of any suppression.  Furthermore, the better frequency coverage of \textsl{Planck} allows for a more reliable exclusion of galactic foreground contamination.}
In appendix \ref{statistics} we give a crude method for estimating the significance, which is in basic agreement with this result. Thus, even assuming that the \textsl{Planck} anomaly will withstand improved analysis of the \textsl{Planck} temperature data, we do {\em not\/} argue that the anomaly is serious enough to require an explanation.

Rather, we have advanced a theoretical argument for a suppression of the CMB spectrum at low $\ell$. If slow-roll inflation was preceded by false vacuum decay, as would be natural in a landscape setting, it becomes more plausible that the \textsl{Planck} anomaly is indicative of a real suppression of the primordial power spectrum.  We will illustrate this quantitatively in Sec.~\ref{models}, by showing how two simple toy models of potential steepening can improve the fit at low $\ell$. We will further argue in Sec.~\ref{sec-probability} that the probability for the onset of such a feature is distributed smoothly over $\log\ell$ with support mainly in the range $1\lesssim \ell\lesssim 10^{4.5}$, rendering a potential onset near $\ell\sim 30$ quite natural.

Given this theoretical motivation, it will be important to corroborate or eliminate any anomaly at large scales.  Can future measurements of cosmological data improve the significance of the \textsl{Planck} anomaly, possibly at the level of discovery?

\subsection{Future Sensitivity}

In estimating the possible sensitivity of future experiments an essential point is that, due to statistical rotational invariance, having access to more modes with the same wavenumber allows a more precise measurement of the power spectrum $P_s(k)$.  Let us briefly review how this applies to the CMB.  One expands the observed temperature anisotropy as
\be
\Delta T(\hat{n})\equiv T(\hat{n})-T_0=\sum_{\ell,m}a_{\ell m} Y_{\ell m}(\hat{n}),
\ee
where $T_0$ is the monopole $T_0=\frac{1}{4\pi} \int d^2\hat{n}T(\hat{n})$.  The $TT$ $C_\ell$'s are then defined as the two-point function
\be
\lan a_{\ell m} a_{\ell'm'}\ran\equiv \delta_{\ell \ell'}\delta_{m,-m'} C_{\ell} ,
\ee
where the average is taken over the ensemble of realizations of the universe. 

 We only get to measure the CMB once, so we cannot observe this average directly.  We can mitigate this limitation by defining an \textit{estimator}
\be\label{eq:estimator}
\hat{C}_\ell\equiv \frac{1}{2\ell+1}\sum_m a_{\ell m} a_{\ell,-m},
\ee
which has the property that on average it is equal to $C_\ell$.  We can evaluate this estimator on the observed set of $a_{\ell m}$'s, and this represents our ``best guess'' for the true $C_{\ell}$.  We can get a sense of the accuracy of this guess by computing the variance
\be\label{oldsig}
\Big\lan\left(\frac{\hat{C}_\ell-C_\ell}{C_\ell}\right)^2\Big\ran=\frac{2}{2\ell+1},
\ee
where we have now assumed that the distribution for the $a_{\ell m}$'s is Gaussian in the sense that higher-point correlation functions can be computed by Wick contraction.  Thus the typical deviation of the estimator $\hat{C}_\ell$ from the true average $C_\ell$ falls like $1/\sqrt{\ell}$.  

We can heuristically understand this as follows: $\hat{C}_\ell$ is roughly an average over $\frac{2\ell+1}{2}$ independently fluctuating quantities $a_{\ell m}a_{\ell ,-m}$, each with mean $C_\ell$ and variance $C_\ell^2$, so the standard deviation of the distribution of values of $\hat{C_\ell}$ falls like $1/\sqrt{\textrm{number of modes}}$, which for the CMB is just $\sim1/\sqrt{l}$.
The estimator in (\ref{eq:estimator}) is optimal, in the sense that no other unbiased estimator can have smaller variance than that.  This \textit{cosmic variance} permanently limits the precision with which the $C_\ell$'s can be determined by CMB observations alone.

However, if different observations gave us access to more modes of a given $\ell$, we could reduce cosmic variance.  We now argue that in fact future measurements of large scale structure (LSS) have the potential to improve the situation considerably.  The key point is that the CMB provides only a two-sphere's worth of information about the primordial power spectrum  $P_s(k)$, so at fixed $\ell\sim k\, D_{ls}$ there are only $\sim \ell$ modes available to average over.   By contrast, the large-scale distribution of galaxies, and perhaps 21 cm radiation, can provide \textit{three-dimensional} information about the power spectrum $P_s(k)$.  For a given $k$, one then has $\sim k^2$ modes to average over.  Compared to the CMB alone, this would decrease the cosmic variance on $P_{s}(k)$ for each measured wavenumber $k$ by an additional factor proportional to $1/\sqrt{\ell}$, where $\ell$ and $k$ are related by $\ell\sim k D_{ls}$. 

Determining the precise cosmic variance for $P_s(k)$ after the inclusion of both the CMB and LSS is beyond the scope of this work, but we can get a rough idea by the following construction.  Imagine that in the future we are able to measure the density of matter inside a cube centered on the earth, with the center of each face at redshift $z$.  The linear size of the cube is $2D_z$, where $D_z$ is determined from equation \eqref{Dls}.  Imposing periodic boundary conditions, the number of modes with wave number between $k$ and $k+dk$, for sufficiently large $k$, is approximately $4\pi \left(\frac{ D_z}{\pi}\right)^3 k^2 dk$.  To make contact with the rest of our paper, we can re-express $k$ in terms of $\ell$ via $\ell=k D_{ls}$, and we can also re-xpress $D_z$ in terms of an effective $\ell$ as $\ell_z\equiv \pi\frac{D_{ls}}{D_z}$.  The total number independent of modes between $\ell$ and $\ell+1$ accessible from the CMB and LSS is then approximately determined by adding the number of modes measured in the CMB to this estimate:
\be\label{modenumber}
N_{\ell}=2\ell+1+4\pi\frac{\ell^2}{\ell_z^3}.
\ee
The new term is correct only when $\ell \gg \ell_z$, but for $\ell_z\gtrsim\sqrt{2\pi}$ this will basically be the case before it begins to compete with the first term.  We can then approximate the new, smaller, cosmic variance of the power spectrum, repackaged as the $C_\ell$'s, as
\be\label{newsig}
\sigma_\ell^2\approx \frac{2}{N_\ell}C_\ell^2
\ee
for all $\ell>\ell_z$ without any large error.  The factor of 2, as in equation \eqref{oldsig}, arises because we are measuring a two-point function that is symmetric under $\vec{k}\to-\vec{k}$.  For $\ell<\ell_z$ LSS does not provide any new information, so we will continue to use the error bars from just the CMB.


In this paper we have in mind a particular, theoretically motivated feature at low $\ell$: a monotonically decreasing suppression of power. It is interesting to ask to what extent including LSS would allow confirmation of such an effect.  As we mentioned in the previous subsection, there is already a hint of such an effect in the current analysis of the \textsl{Planck} data, and in the following section we will present two models which improve the fit to the current data.  There is a simple way to estimate how sharply future LSS experiments could distinguish these models from $\Lambda$CDM.  Suppose that the  second of our models, the power-law model of section \ref{sec-inverse}, is correct.  We can then generate future ``data'' for the $C_\ell$'s independently for each $\ell$, from a Gaussian ensemble centered on this model. The standard deviations are taken as \eqref{newsig} for $\ell>\ell_z$ and \eqref{oldsig} for $\ell<\ell_z$.   Remember that here $\ell_z$ should be thought of as setting the largest scale out to which we can measure 3D information from LSS.  

\begin{figure}
\begin{centering}
\includegraphics[height=4.2cm]{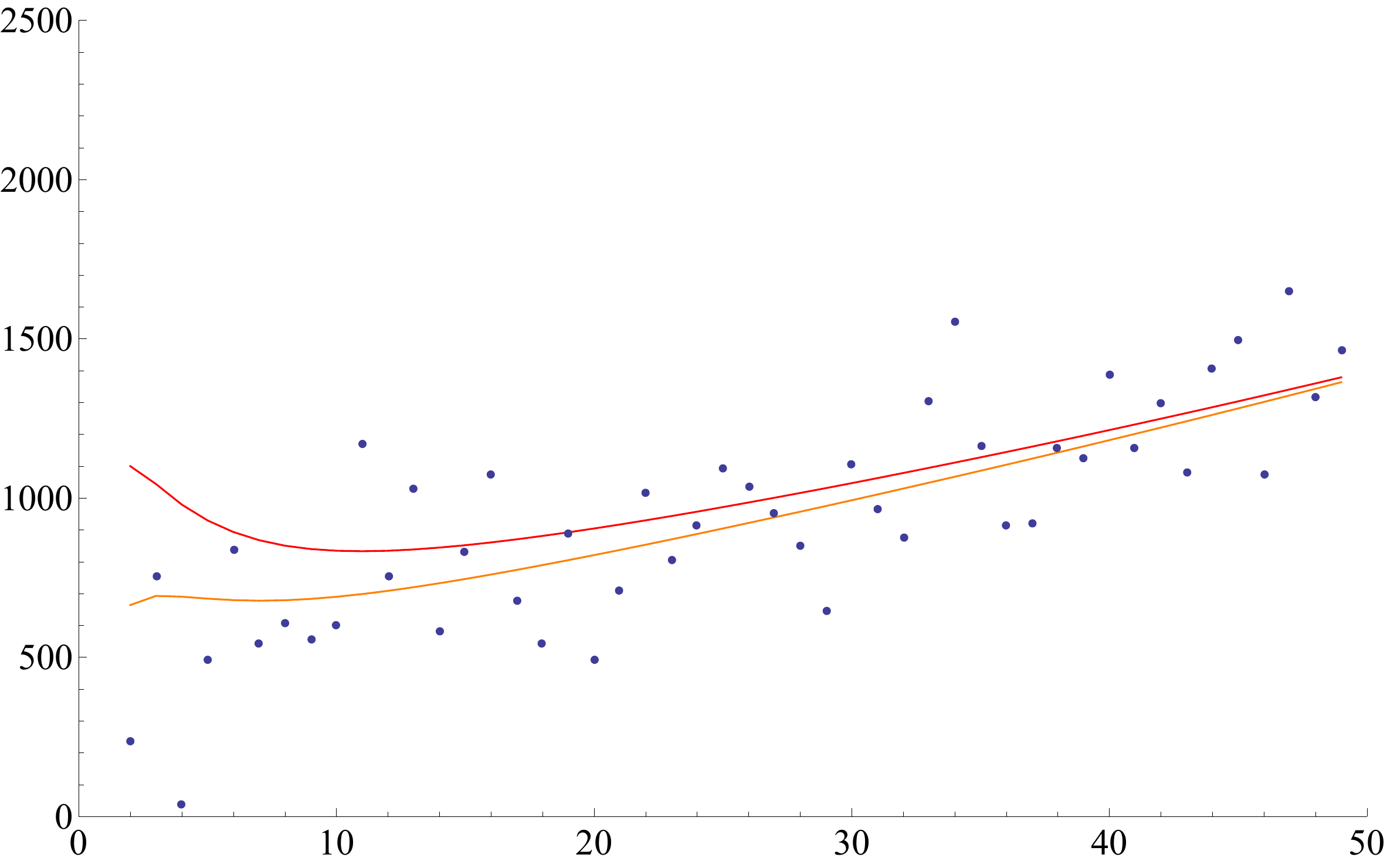}\hspace{.7cm}\includegraphics[height=4.2cm]{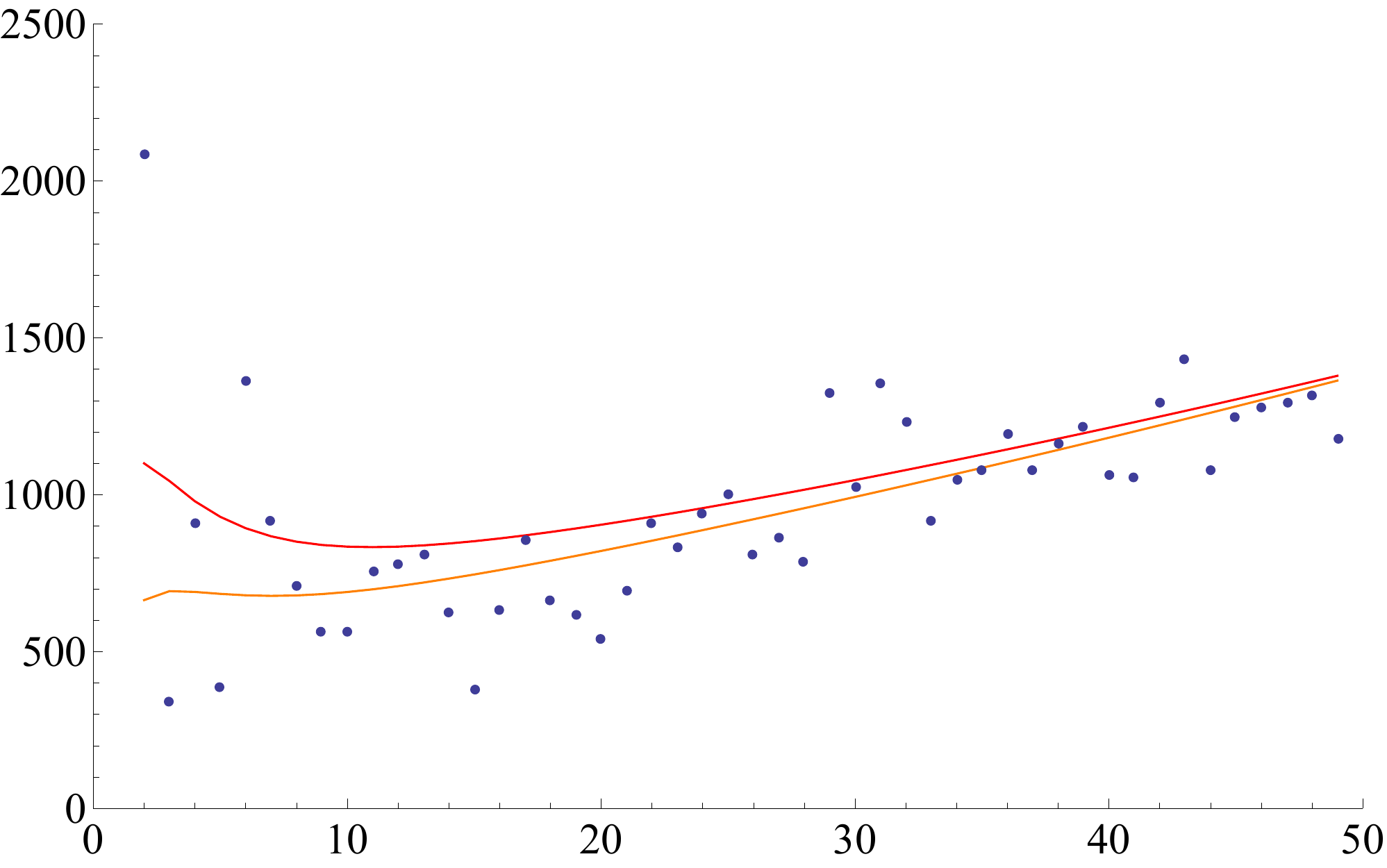}
\caption{``Data'' for the $\frac{\ell(\ell+1)}{2\pi}C_\ell$'s generated from the power-law model.  The red curve is $\Lambda$CDM, while the green curve is the model we introduce in section \ref{sec-inverse}.  On the left we include no measurements of large scale structure, while on the right we include LSS measurements only down to $\ell_z=6.7$.  The ``data'' on the left differs from $\Lambda$CDM at $2.4\sigma$, while that on the right differs at $3.4\sigma$.}\label{fake1}
\end{centering}
\end{figure}

\begin{figure}
\begin{centering}
\includegraphics[height=4.2cm]{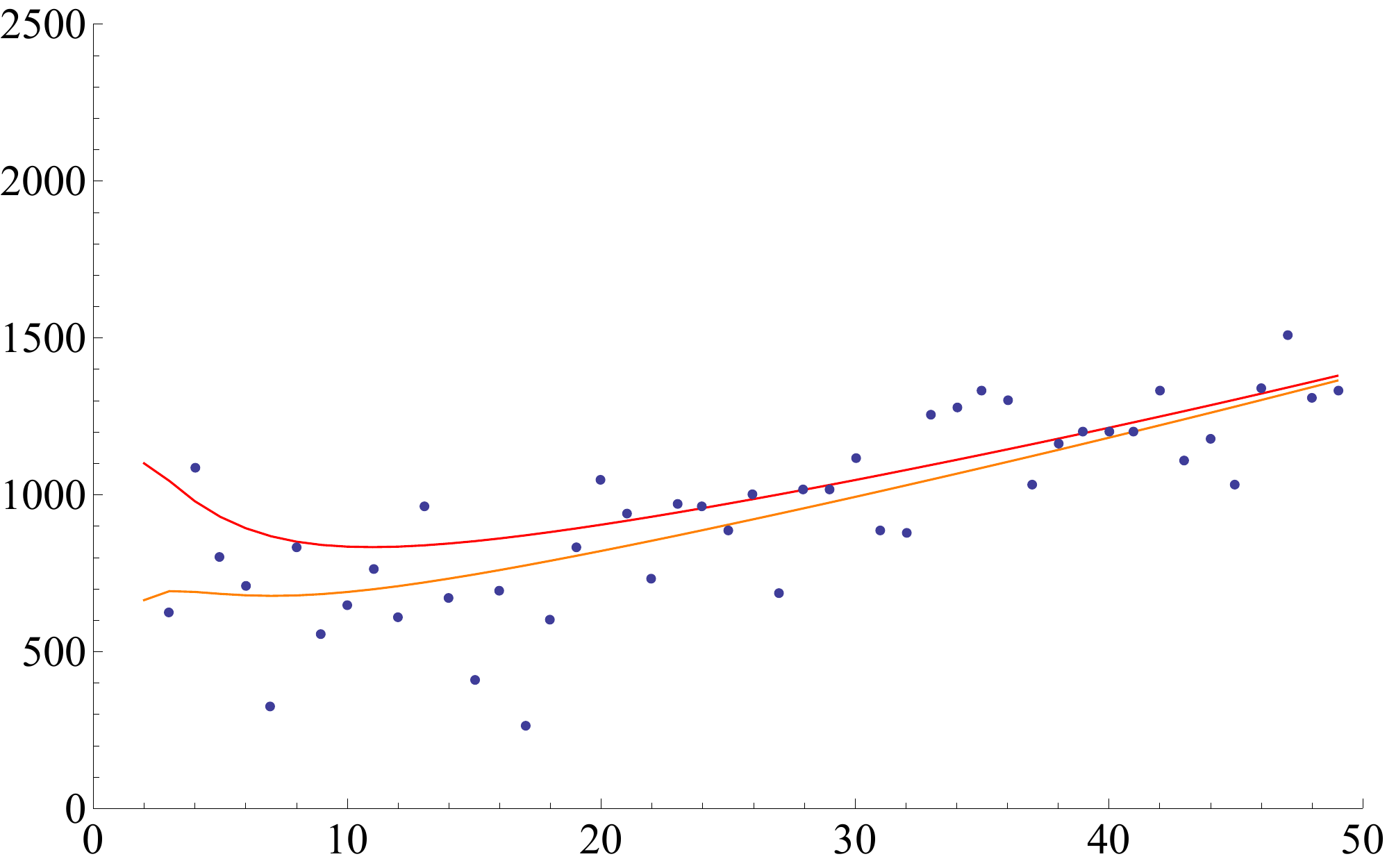}\hspace{.7cm}\includegraphics[height=4.2cm]{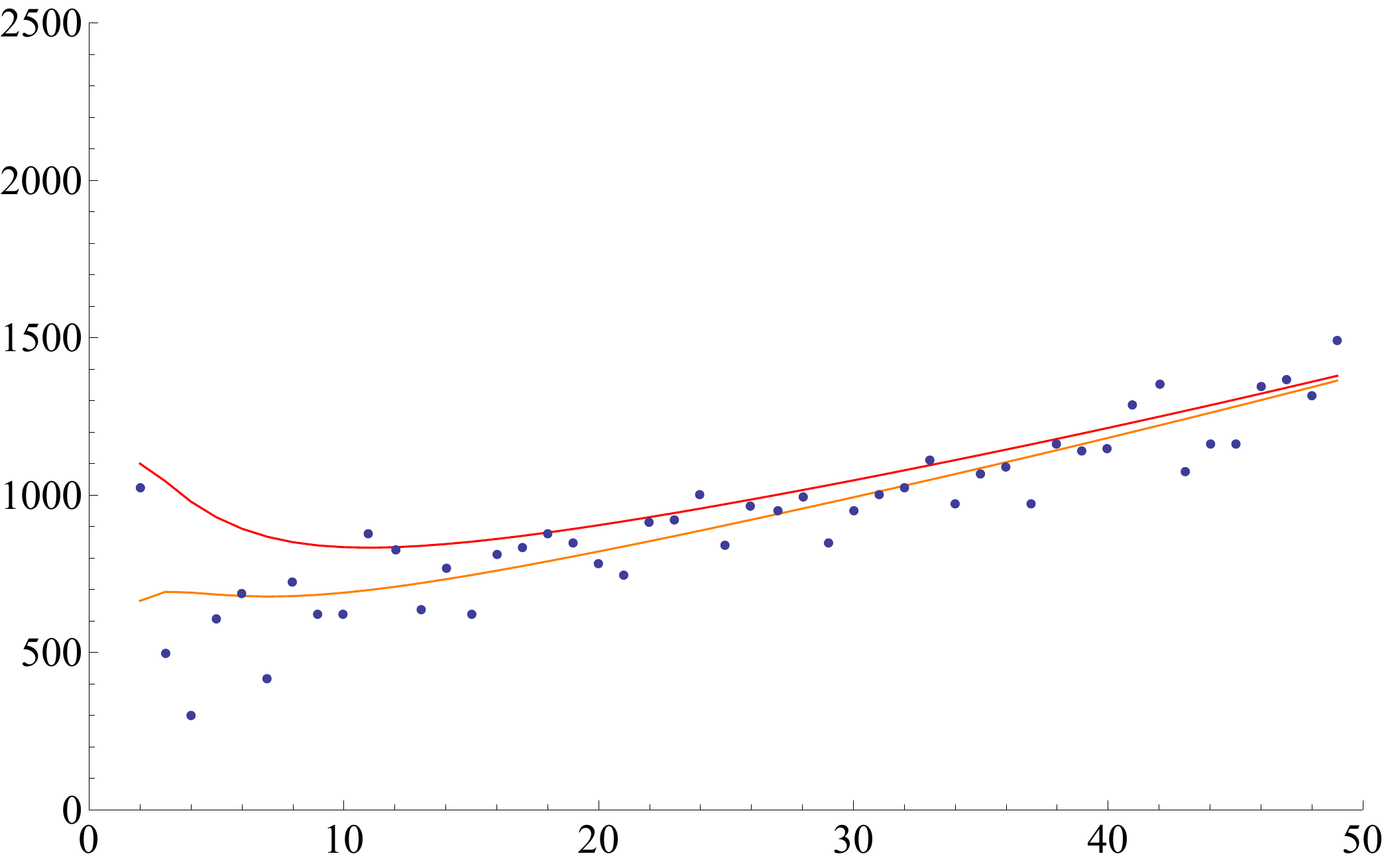}
\caption{More ``data'' from the power-law model.  On the left LSS is included down to $\ell_z=5.5$, while on the right it is included down to $\ell_z=3.6$.  The left side ``data'' differs from $\Lambda$CDM at $3.8\sigma$, while the right differs at $5.2\sigma$.}\label{fake2}
\end{centering}
\end{figure}

For $\ell_z=50$ the ``data'' we generate should resemble figure \ref{datafig2}, while for lower values of $\ell_z$ it should move closer to our exponential model. We illustrate this in figures \ref{fake1}, \ref{fake2}, where we plot some fairly typical instances of ``data''  for $\ell_z =50, 6.7, 5.5$, and~$3.6$.  Finally, we can use our crude statistical techniques from appendix \ref{statistics} to estimate the probability that this ``data'' arises from $\Lambda$CDM, again with the error bars appropriately reduced thanks to the LSS ``data''.  To focus on the low~$\ell$ region we use the diagnostic with $\ell_{max}=40$.


We have chosen the numbers $6.7, 5.5,$ and $3.6$ for $\ell_z$ because the first two are roughly what might be expected from upcoming EUCLID-like galaxy surveys, while the third is what might come out of 21 cm measurements.  More explicitly, $\ell_z=6.7$ corresponds to $z\approx 3$, while $\ell_z=5.5$ corresponds to $z\approx 5$. If we imagine that 21 cm measurements can map gas out to $z\approx 50$ then this gets us down to $\ell_z=3.6$.  Generating the data 20-30 times for each choice of $\ell_z$, we find that with $\ell_z=50$ the significance of the deviation from $\Lambda$CDM is typically of order $2-3\sigma$.
This is in rough agreement with what \textsl{Planck} has found.  Including LSS with $\ell_z=6.7$, the typical significance increases to about $3-4\sigma$.
For $\ell_z=5.5$ it increases to about $3.5-4.5\sigma$,
while for $\ell_z=3.6$ it is typically in the vicinity of $5-6\sigma$.

Thus, we find that future LSS surveys may have the statistical power to strongly corroborate or exclude a low-$\ell$ power suppression of the type suggested by the Planck anomaly. Our statistical methods have admittedly been rather rough, but a more precise analysis that takes into account the details of the experiments is quite possible.  Our significance results are rather sensitive to the order one coefficient in front of $\frac{\ell^2}{\ell_z^3}$ in equation \eqref{modenumber}, so it will be very important to compute this more carefully. Of course, in an actual LSS survey, there will also be systematic challenges to face in order to approach these asymptotic estimates. 

Although we have focused on a single model, this analysis shows that future LSS measurements may in general allow strong constraints on models that predict deviation from $\Lambda$CDM at low $\ell$.  If the actual inflationary potential differs from LambdaCDM more than our model, future observations could rule out $\Lambda$CDM at even greater levels of confidence.  We hope we have made it clear how important this effort might be, and the potential discoveries it could lead to.

Finally, let us comment on polarization measurements.  Given that $E$ and $T$ modes are quite uncorrelated, due to difference in the visibility function, measurement of the $EE$ power spectrum will also give us access to more independent modes, in fact twice as many in the limit where $T$ and $E$ are completely uncorrelated.  However, some of the difference in the visibility function comes from the contribution to polarization from the epoch of reionization. This unfortunately contaminates the low-$\ell$ signal with contribution from higher wavenumbers than the ones contributing to the same $\ell$'s at recombination.  Removing this effect would increase the $TE$ correlation, decreasing the potential improvement to the statistical significance of the \textsl{Planck} anomaly from $EE$ measurements, so we expect the benefit of these measurements to be limited, although non-negligible, and we do not study them in detail here (for a discussion of many of the relevant issues see \cite{Mortonson:2009qv}).\footnote{To get a rough idea of size, if we imagine we get $4/3$ as many independent modes from polarization this should allow us to multiply the existing significance from the $TT$ anomaly by $\sqrt{4/3}\approx1.15$.  We then might expect increases of order a few tenths of a $\sigma$.}  As we already discussed however, the potential steepening does lead to power suppression directly in the $TE$ and $EE$ CMB, and this should eventually be detectable with at least some significance.  More optimistically, if $BB$ is ever measured than eventually it might be possible to see that it is \textit{not} suppressed at low $\ell$, as predicted in section \ref{pertsec}.

\section{Two Models}
\label{models}

We will now illustrate our general discussion with two simple models. In the first, the steepening of the potential is rapid enough that the CMB power suppression is closely related to the onset of inflation, when $\Omega_K\simeq1$. As expected, one finds that this model is already fairly tightly constrained; in models of this type, a future discovery of curvature is likely. The second model illustrates a more gradual steepening feature which allows parametric separation of the onset of CMB power suppression and the beginning of inflation. It is thus much less constrained. In this model, observation of the CMB suppression is possible even if curvature is never found.  We show that both models are able to improve the fit to the \textsl{Planck} data at low $\ell$. 

Both models are toy models. They represent the two extreme possibilities for the suppression of the curvature while exhibiting power suppression at low~$\ell$. We expect that realistic inflationary models with these characteristics can be constructed, but we will not attempt this here.
 
\subsection{Exponential Steepening}
\label{sec-exp}

\begin{figure}
\begin{center}
{\includegraphics[height=6cm]{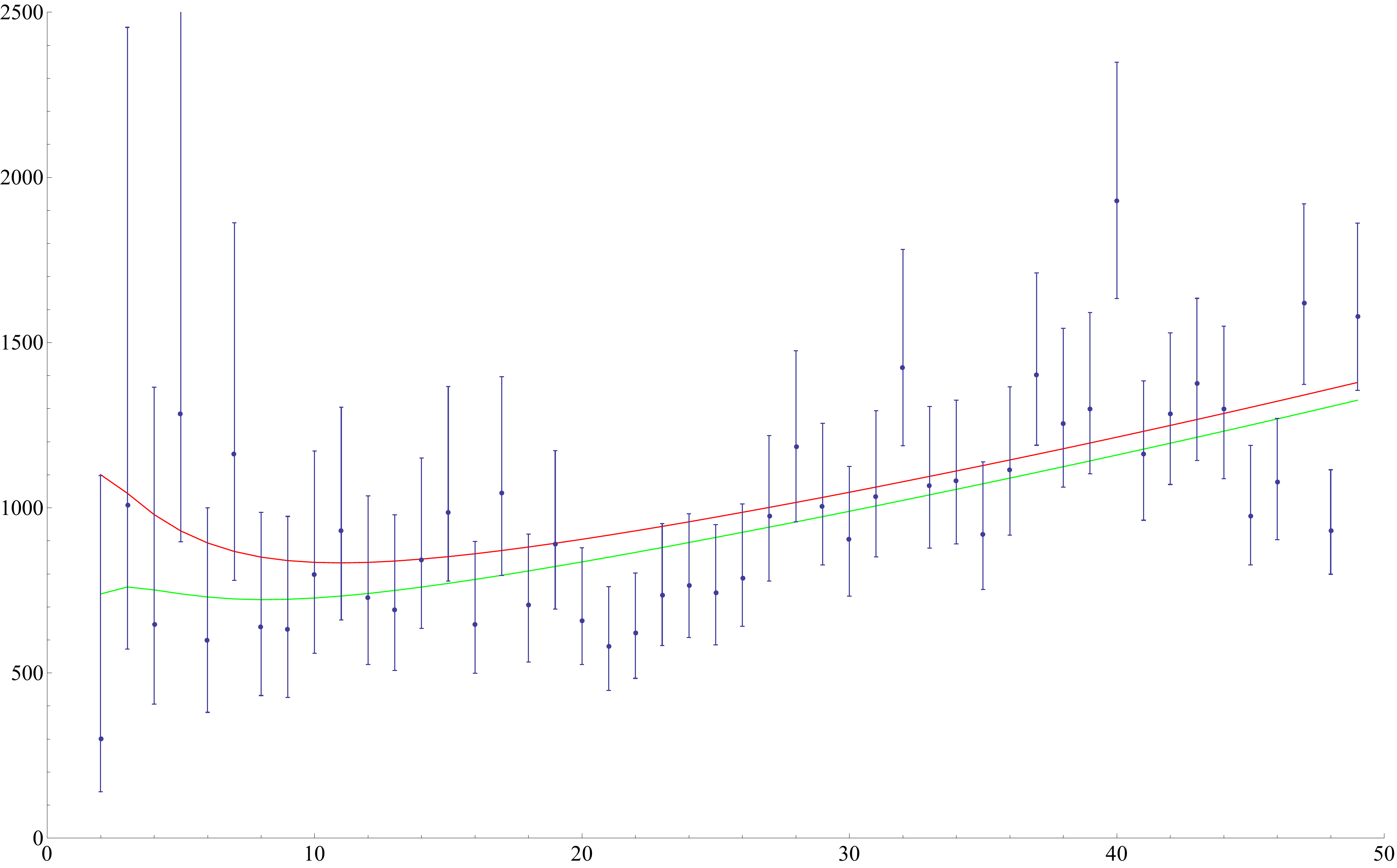}}
{\includegraphics[height=6cm]{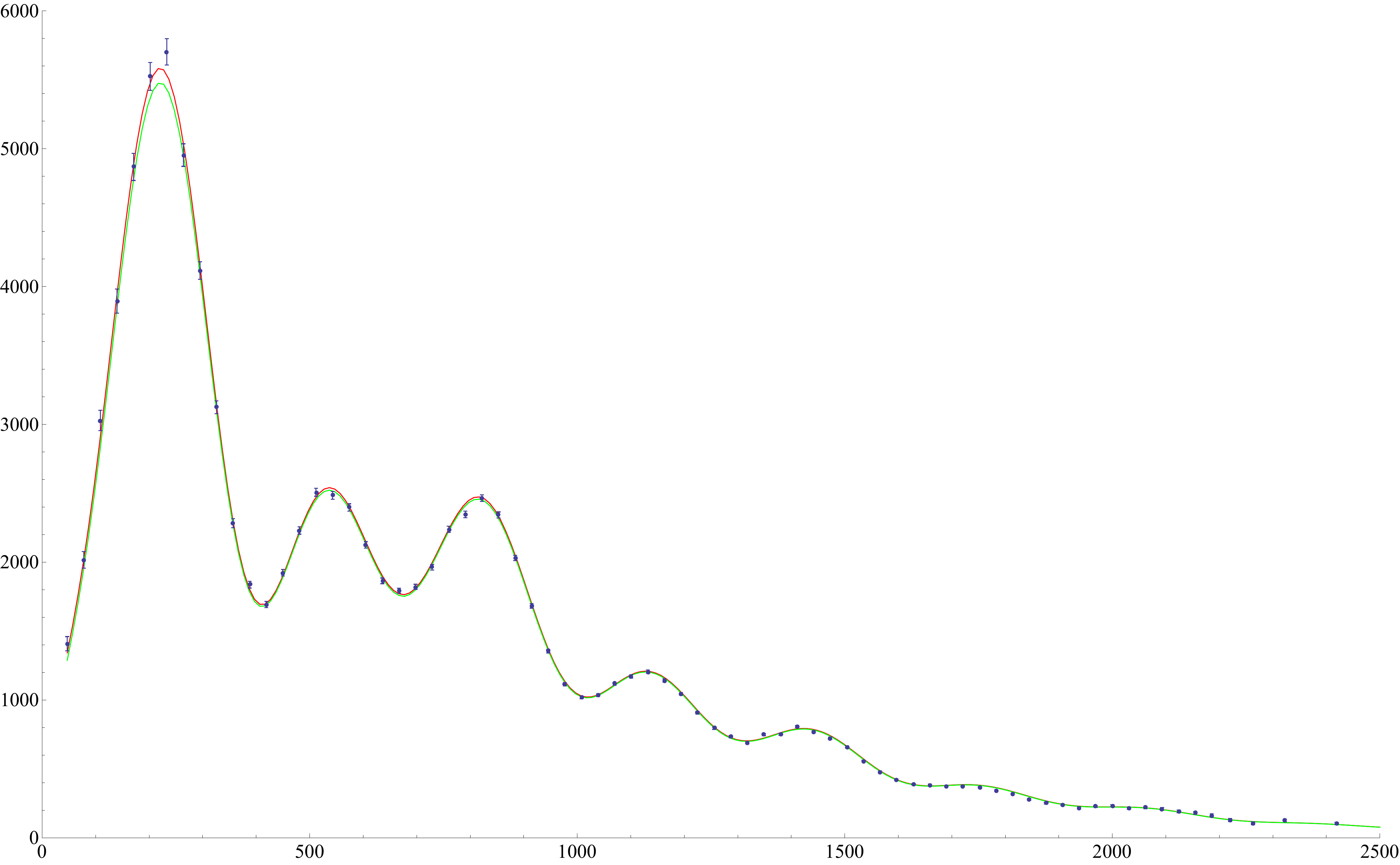}}
\end{center}
\caption{Comparison of $\frac{\ell(\ell+1)}{2\pi}C_\ell$ for the model \eqref{expP} in green, $\Lambda$CDM in red, and the \textsl{Planck} data.  Theory curves are computed using CLASS \cite{Lesgourgues:2011re,Blas:2011rf,Lesgourgues:2011rg}.}\label{expplot}
\end{figure}

We first take 
\begin{align}\nonumber
V_S&=V_i\left(1-\sqrt{2\epsilon_S}\frac{\phi}{M_P}\right)\\
V_R&=V_i e^{-\frac{\phi}{M}}.\label{expmodel}
\end{align}
This model has two parameters, $\gamma$ and $M$.  It is convenient to define
\be
n\equiv \sqrt{2\epsilon_S}\frac{M_P}{M},
\ee
and instead work in terms of $\gamma$ and $n$.  Equations \eqref{phik}, \eqref{spectrares} then give
\be\label{expP}
\mathcal{P}_s(k)=\mathcal{P}_{s,S}(k)\left[1-\gamma\frac{n}{\epsilon_S}\left(\frac{k_h}{k}\right)^n+\ldots\right] .
\ee
We need to take $n\sim 1$ in order to fit the low $\ell$ power suppression in Planck. With this constraint, we need to take $\gamma$ small enough so that the correction to $\epsilon_S$ is still perturbative for $\phi>0$.
This model is plotted against the data in Fig.~\ref{expplot} for $n=.7$ and $\gamma=3.5\times10^{-3}$.\footnote{This plot is slightly dishonest for the following reason; the simple $V_S$ from equation \eqref{expmodel} was chosen to make our discussion simple, but strictly speaking it does not quite match $\Lambda$CDM with the \textsl{Planck} parameters since it predicts $r\approx .1$ which is a bit large.  This has the effect of slightly enhancing the $C_\ell$'s at low $\ell$, by of order a few percent, so to be consistent with \textsl{Planck} in Fig.~\ref{expplot} we have set $r=0$ for both curves.  This can be justified by slightly modifying $V_{S}$ by making the slow roll parameter $\eta$ non-vanishing, and therefore to suppress $r$, but since our models are really just heuristic illustrations anyway we have not done so.  This comment also applies to Fig.~\ref{powplot} below.}
Using the crude statistical measure from the appendix, we find that for $\ell_{max}=30$ the significance of the low-$\ell$ anomaly is decreased from  about $2.4\sigma$ to  $1.0\sigma$, while for $\ell_{max}=49$ it decreases from  $1.8\sigma$ to $.36\sigma$.  From the figure it seems that we could fit even better by increasing both $\gamma$ and $n$ a little bit, but as we now argue the exponential growth of this potential allows so little inflation for $\phi<0$ that doing so would produce observable curvature.  

\begin{figure}
\begin{center}
{\includegraphics[height=4cm]{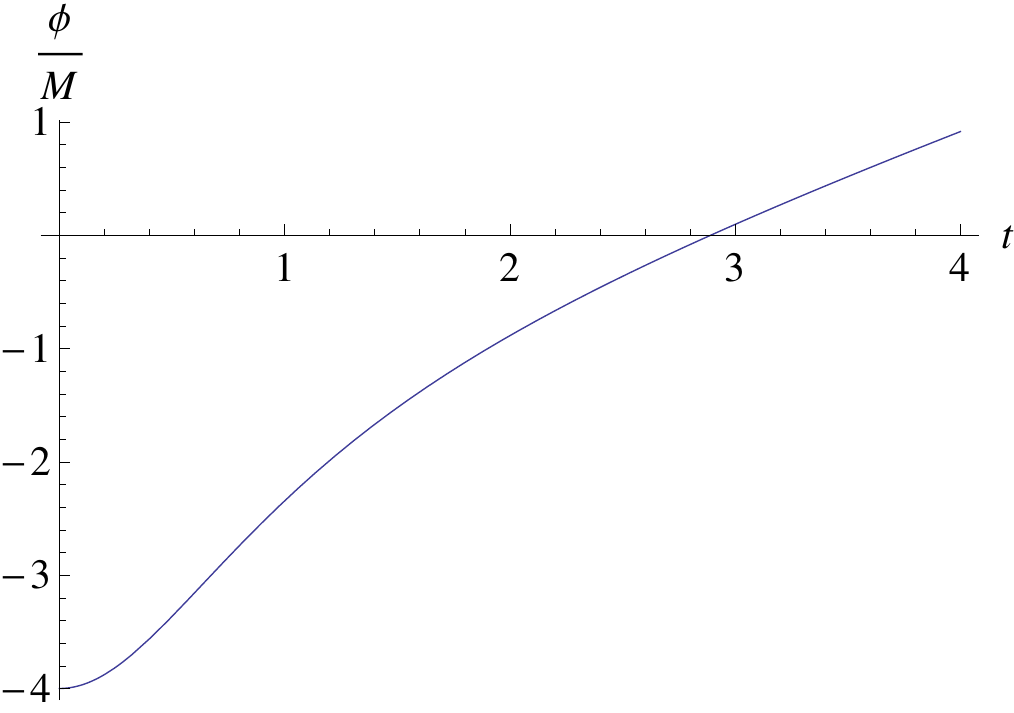}}\hspace{.1cm}
{\includegraphics[height=4cm]{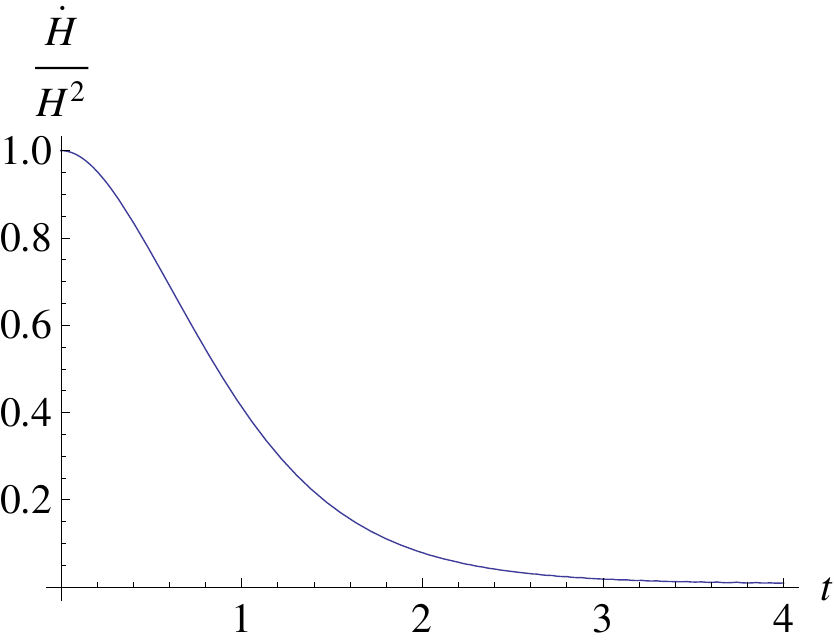}}
\end{center}
\caption{Plots of $\phi/M$ and $\frac{\dot{H}}{H^2}$ from a numerical solution of \eqref{frw}.  The field is dropped at $\phi/M=-4$, and the time is measured in units of the Hubble constant on the plateau, $\sqrt{\frac{V_i}{3M_p^2}}$.  The second plot shows that inflation does not really begin until soon before horizon crossing at $t\approx 2.9$.}\label{numsol}
\end{figure}

\begin{figure}
\begin{center}
\includegraphics[height=5cm]{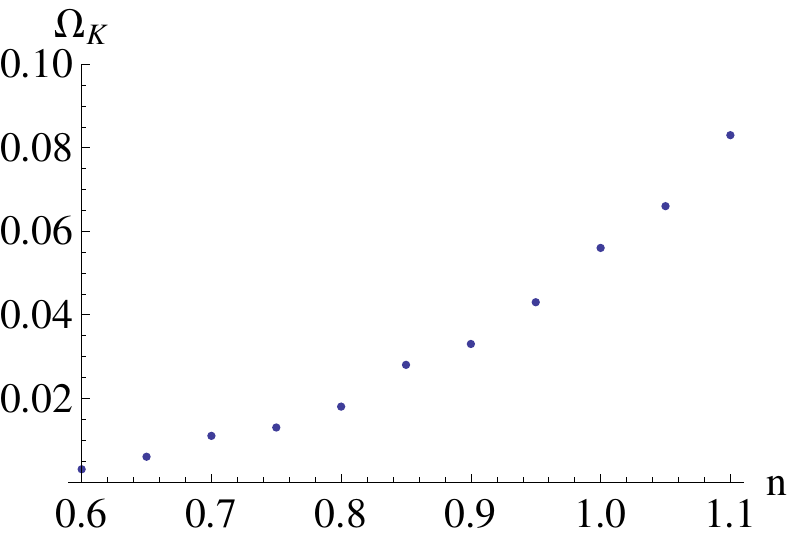}
\end{center}
\caption{$\Omega_K$ as a function of $n$.  A rough analytic treatment suggests approximating this function as $e^{-\frac{2.5}{n}}$.}\label{omegakplot}
\end{figure}

When $\phi<0$ in this model, the potential rapidly approaches $V_i\gamma e^{-\frac{\phi}{M}}$.  When this happens, the slow-roll parameters are both proportional to $\left(\frac{M_P}{M}\right)^2=\frac{n^2}{2\epsilon_S}$, so the slow roll approximation fails almost immediately.  To compute the size of curvature in the model we need to solve the full FRW equations
\begin{align}\nonumber
H^2&=\frac{1}{a^2}+\frac{1}{3M_P^2}\left[\frac{1}{2}\dot{\phi}^2+V(\phi)\right]\\\label{frw}
\ddot{\phi}&=-3H\dot{\phi}-V'(\phi),
\end{align}
and then evaluate $\Omega_K=(a_hH_h)^{-2}$.  General properties of the solutions of these equations are discussed in detail for example in \cite{Dong:2011gx}; the initial conditions from the CDL instanton \cite{Coleman:1980aw} are that $\dot{\phi}=0$, $a=0$, $\dot{a}=1$, and $\phi$ is some negative number such that $|\phi|$ is significantly greater than $M$ but $V(\phi)\ll M_P^4$.  $\gamma$ and $n$ are taken from the quoted values above, and we take $V_i=3 \times 10^{-9} M_p^4$ and $\epsilon_S=0.006$ to match the \textsl{Planck} data using equations \eqref{Ps}, \eqref{ns}.  A rough analytic treatment is possible, but it is more convenient to simply solve them numerically.  For the quoted values of parameters this gives $\Omega_K=0.011$, which is just barely consistent with the current $2\sigma$ bound, $-0.028<\Omega_K<0.008$, from (\textsl{Planck}+lensing+WP+highL) \cite{Ade:2013lta}.  Some plots of the solution are shown in Fig.~\ref{numsol}.  We show some values of $\Omega_K$ as a function of $n$, with $\gamma$ chosen to approximately fit the \textsl{Planck} anomaly, in Fig.~\ref{omegakplot}.

The simplicity of the exponential potential in this model produces an interesting tension between curvature and cosmic variance.  From Fig.~\ref{omegakplot} we see that getting enough inflation to dilute curvature requires $n<1$, but since the deviation of the power spectrum from $\Lambda$CDM falls off like $\frac{1}{\ell^n}$ at large $\ell$ we need $n>.5$ for this to fall off faster than cosmic variance.  Above we chose $n=.7$, which satisfies both of these constraints, but not by much.  This tension can be relaxed either by allowing $V_R$ to fall off faster at large $\phi$ or making it a little less steep for $\phi<0$.
 
\subsection{Power Law Steepening}
\label{sec-inverse}

\begin{figure}
\begin{center}
\includegraphics[height=6.5cm]{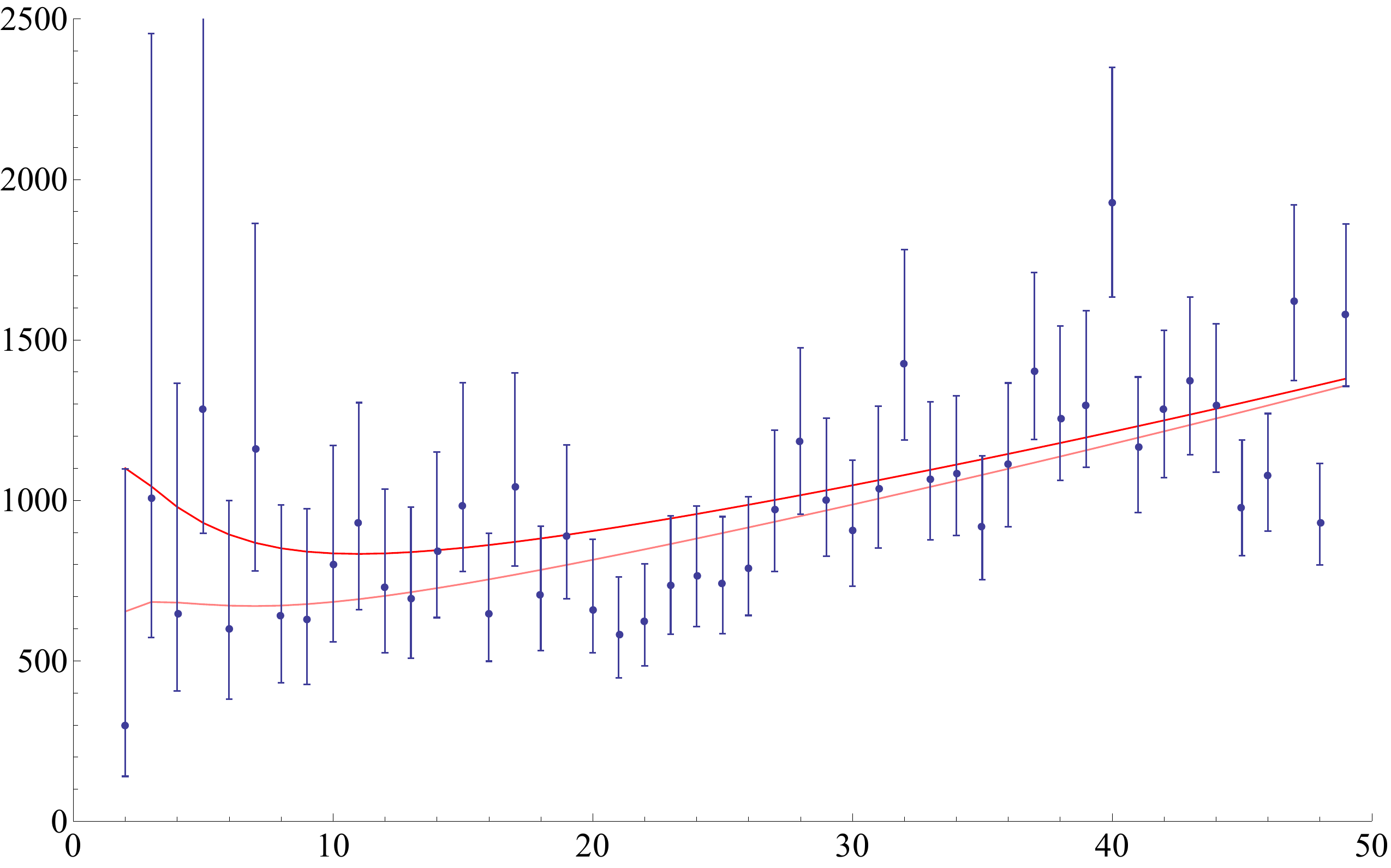}
\end{center}
\caption{Comparison of $\frac{\ell(\ell+1)}{2\pi}C_\ell$ for the model \eqref{powmol} in orange, $\Lambda$CDM in red, and the \textsl{Planck} data.}\label{powplot}
\end{figure}

The previous model was an example of a potential that steepens rapidly enough that it is very difficult to have an observable suppression of power without observable curvature.  We now consider a more gentle model, with the same $V_S$ as before but now with
\be\label{powmol}
V_R=\Theta(\phi_{c}-\phi)\,\frac{V_i}{\varsigma}\left(\frac{\phi_{c}-\phi}{M_P}\right)^\varsigma \ ,  \quad{\rm with}\quad {\varsigma>1} \ .
\ee
Here $\Theta$ is the Heaviside theta function; this potential is non-analytic at $\phi_{c}$, but that is no issue for the illustrative purposes of our toy model.  For future convenience we parametrize $\phi_{c}$ as
\be
\phi_{c}\equiv \sqrt{2\epsilon_S}\log \frac{\ell_{c}}{D_{ls}H_0}.
\ee
Equation \eqref{spectrares}, at leading order in slow roll and $\gamma$, then gives
\be
\mathcal{P}_s(k)=\mathcal{P}_{s,S}(k)\left[
1-\Theta\left(\frac{\ell_c}{D_{ls}}-k\right)\frac{2\gamma}{\sqrt{2\epsilon_S}}\left(\sqrt{2\epsilon}\log\frac{\ell_c}{D_{ls} k}\right)^{\varsigma-1}+\ldots \right]
\ee
The correction turns off for $k D_{ls} >\ell_c$, which in the CMB corresponds to $\ell>\ell_c$.  It is quite easy to fit the data by varying $\varsigma$, $\gamma$, and $\ell_{c}$.  For example for $\varsigma=2.3$, $\gamma=.08$, and $\ell_{c}=65$ using the method from the appendix we find the significance decreases from $2.4\sigma$ to $.29\sigma$ for $\ell_{max}=30$ and from  $1.8\sigma$ to  $.35\sigma$ for $\ell_{max}=49$.  A comparison of this model to standard $\Lambda$CDM and the data is shown in Fig.~\ref{powplot}.
It may appear that the improvement of fit for this model compared to the previous one came from having three parameters as opposed to two, but $\ell_c$ does not really have too much effect on the curve in the relevant region.  The important difference is that in this model it is easy to inflate away the curvature, so we are free to take a larger value for $\gamma$ without running into the observational bound on $\Omega_K$.  

To see that this model indeed allows inflation for $\phi<0$, we observe in this region the potential is essentially just $\sim\gamma V_i \left(\tfrac{\phi}{M_P}\right)^\varsigma$.  The slow roll parameters are both proportional to $\left(\tfrac{M_P}{\phi}\right)^2$, which are quite small for $|\phi|\gg M_P$.  The number of e-foldings prior to $\phi=0$ will thus be approximately $\left(\tfrac{\phi_i}{M_P}\right)^2$, where $\phi_i$ is the point where curvature domination ends. The number of e-foldings in the past is limited by the onset of slow roll eternal inflation, as in that case we have $k^3 \mathcal{P}_s(k)\sim\Omega_K\sim 1$, and so, having inflation in that regime does not help diluting curvature anymore.\footnote{ Slow-roll eternal inflation is a situation where the scalar field is rolling slowly enough that quantum fluctuations carry it up the potential as often as classical rolling takes it down \cite{Linde:1986fd}.  In this situation, inflation lasts forever globally, but in each region of space sooner or later by chance the field fluctuates down the potential enough that it then rolls smoothly down. This allows a fairly homogeneous inflating region to be created in a manner that is crudely similar to bubble nucleation. Instead modes that exited the horizon during the eternal inflation period induce density perturbations $H^2/\dot{\phi}$ of order one. While for many years a quantitive and sharp understanding of slow roll eternal inflation was lacking, recently relevant progress in this direction has been made~\cite{Creminelli:2008es,Dubovsky:2008rf,Dubovsky:2011uy}} To avoid slow roll eternal inflation we need $\gamma V_i\left(\tfrac{\phi_i}{M_P}\right)^{\varsigma+2}\ll M_P^4$, so we can get at most a number of e-foldings before horizon entry of order
\be
\mathcal{N}\approx \left(\frac{M_P^4}{\gamma V_i}\right)^{\frac{2}{\varsigma+2}}\ .
\ee
This can be quite large.  We thus view this model as an extreme case where curvature and steepening can be separated almost arbitrarily.

\section{Statistical Tuning in a Large Landscape\label{sec:probpotential}}
\label{gensec}

In the former sections, we described predictions for observables, {\it given} a certain potential. We discussed how to observationally recognise an inflationary potential with a steepening at large scales, as in Fig.~\ref{potentialfig}.  We argued qualitatively that such potentials are natural if slow-roll inflation followed the decay of a false vacuum. 

In this and the next section, we will turn to a quantitative question: what is the {\it likelihood} that our own inflationary history contains a potential with steepening, and what is the probability that this feature is observable? This question can only be posed if many possible potentials exist, that is, if there exists a large landscape of effective potentials. In order to answer it, we would need to compute the statistical distribution of potentials. For the landscape of string theory, this task lies beyond our current understanding. However, we can make progress by considering a range of plausible distributions and investigating their implications.

We stress that the conclusions of the previous sections are not affected by the following discussion. They do not depend on the assumptions we are about to make.

%
\subsection{The Landscape of String Theory}

The landscape of string theory provides a setting in which low energy parameters can take on many different values. Their statistical distribution is controlled by the underlying unique fundamental theory. If this prior distribution can be computed or at least constrained, probability distributions over observed values of parameters can be obtained by conditioning on observers. This allows for quantitative and falsifiable predictions. At present, the landscape provides the only viable explanation of the smallness of the cosmological constant~\cite{Sak84,Bousso:2000xa}. The landscape can also explain other fine-tuning or coincidence problems.  The hierarchy problem increasingly appears to be of this type, as no evidence for naturalness of the weak scale has so far been found at the LHC or by any other experiment. 

Any theory that contains at least one long-lived metastable vacuum with positive cosmological constant (such as, apparently, our own~\cite{Per98,Rie98}) leads to eternal inflation: globally, the universe grows faster than it decays. Thermal effects and vacuum decays give rise to infinite recurrences and obstruct the computation of relative probabilities. The question of how to regulate these infinities is known as the measure problem of eternal inflation; see, e.g., Ref.~\cite{Guth:2007ng,Freivogel:2011eg} for reviews. Any viable proposal must reproduce the standard probabilities for the outcome of laboratory experiments and closely related physical processes.\footnote{This requirement turns out to be a powerful constraint (which is fortunate, since no fundamental derivation of the correct measure is known). Measure proposals that survive this constraint have proven remarkably successful phenomenologically. In particular, the causal patch measure allows for a direct explanation of the Why Now coincidence~\cite{BouHar07} and other cosmological coincidences~\cite{Freivogel:2008qc,BouHal09,BouHar10,Bousso:2013rda}. This has improved on the predictions of classic arguments (e.g.~\cite{Linde:1984ir,Banks:1984cw,Weinberg:1987dv,Vilenkin:1996ar,Tegmark:1997in,TegAgu05}) quantitatively, while eliminating their specific anthropic assumptions.  A small set of closely related, phenomenologically interesting proposals also remain viable~\cite{Bou06,DeSimone:2008bq,Salem:2011mj}.}  Perturbative effects on the CMB spectrum are of this type~\cite{Salem:2012ve}, so we will be able to ignore the measure problem for the purposes of our analysis.

The main idea that will go into our discussion of the genericity of potentials is that having flat directions is \textit{statistically tuned}.  By statistically tuned we do \textit{not} mean that it is radiatively unstable in the way that the Higgs boson mass is (without a new symmetry at the weak scale).  A potential which is approximately flat at tree level will also be flat at higher order in perturbation theory, since in the limit of zero slope there is a shift symmetry which emerges.  This type of argument cannot explain however why the tree-level potential was flat in the first place,  and it is reasonable to ask  what type of tree-level potentials are typical. 

The question of genericity of potentials goes beyond the usual particle-physics notion of tuning. But in a large landscape, the question is well-defined. Its answer depends on the statistical distribution of potentials, which can at least in principle be derived from the underlying theory. To emphasize the distinction from the standard notion of tuning, we will refer to the atypicality of a feature in the landscape as {\em statistical tuning}. Atypicality in the prior distribution need not conflict with observation. But if an observed feature is statistically tuned even after conditioning on observers, then it rules out the theory at the corresponding level of confidence.  


\subsection{Steepening vs.\ Flattening}

As a first application of this framework, we will now motivate our assumption that our region of the universe is inside a bubble nucleated in a metastable false vacuum.  Why could the era of standard slow-roll inflation not instead be preceded by slow-roll eternal inflation?  In this situation there would be no potential barrier of the type shown in Fig.~\ref{potentialfig}, so there would be no generic argument for a steepening feature.  We suggest however that this type of eternal inflation requires significantly more statistical tuning of the potential than the false vacuum eternal inflation we have been considering so far.  Consider an extremum in a multi-field landscape, where the potential obeys
\be
\frac{\partial V}{\partial \phi_n}=0 \qquad \forall n.
\ee
A point where either false-vacuum or slow-roll eternal inflation is happening should satisfy this to a very good approximation.  Let us now consider the eigenvalues $m^2_n$ of the second-derivative matrix $\frac{\partial^2 V}{\partial \phi_n \partial \phi_m}$.  In order to be able to neglect quantum gravity and in order to have inflation, we roughly need
\be\label {m2bounds}
-\frac{H^2}{M_p^2}=-\frac{V}{3M_p^4}<\frac{m^2_n}{M_p^2}<1 \qquad \forall n.
\ee
The first inequality ensures that the field will not roll down a tachyonic direction before inflating, and is equivalent to $\eta>-1$.  The second inequality ensures that the masses are not Planckian.  The point however is that the lower bound on each $m^2_n$ is much smaller in absolute value than the upper bound, so typically we can expect all of the eigenvalues to be positive and significantly larger than $H^2$.  In this case we have a false-vacuum, which decays via bubble nucleation as in Fig.~\ref{potentialfig}.  More quantitatively, let's imagine that the energy scale of the potential is of order $10^{-1}$ in Planck units.  We get to raise this to the fourth power in equation \eqref{m2bounds}, so if we imagine a uniform distribution for each $m^2$ over the allowed range then the probability is roughly $10^{-4}$ per direction that we get slow-roll eternal inflation.  This is not quite correct, since because of eigenvalue repulsion the different eigenvalues are not all independent.  This problem has been studied quantitatively for the Gaussian matrix ensemble in \cite{Marsh:2013qca}, and applying their results we find that for $\left(\frac{V}{M_p}\right)^{1/4}\sim 10^{-1}$ we need at least of order $10^2$ fields to have a decent chance of even a single tachyonic direction.\footnote{We thank Timm Wrase for help in understanding and using their results.}  More realistically we should probably demand the energy be even lower in Planck units, which causes this number to increase quadratically in the energy scale.  We emphasize that we are \textit{not} claiming that it is impossible to have slow-roll eternal inflation somewhere in the landscape, but only that we can argue that our immediate ancestor is most likely to be a false vacuum.

This argument can also be applied to the question of whether we should have included other field directions near the inflationary plateau in Fig.~\ref{potentialfig}.  In fact, multifield inflation is even more disfavored from this point of view, since during the observable period of inflation, $\frac{V}{M_p^4}\lesssim 10^{-12}$. (Of course, even if there were other fields whose masses are comparable to Hubble on the inflationary plateau of Fig.~\ref{potentialfig}, one would still expect the field trajectory to steepen near a potential barrier.)

\subsection{Coleman-De Luccia vs.~Hawking-Moss Decay}

\begin{figure}
\begin{center}
\includegraphics[height=4cm]{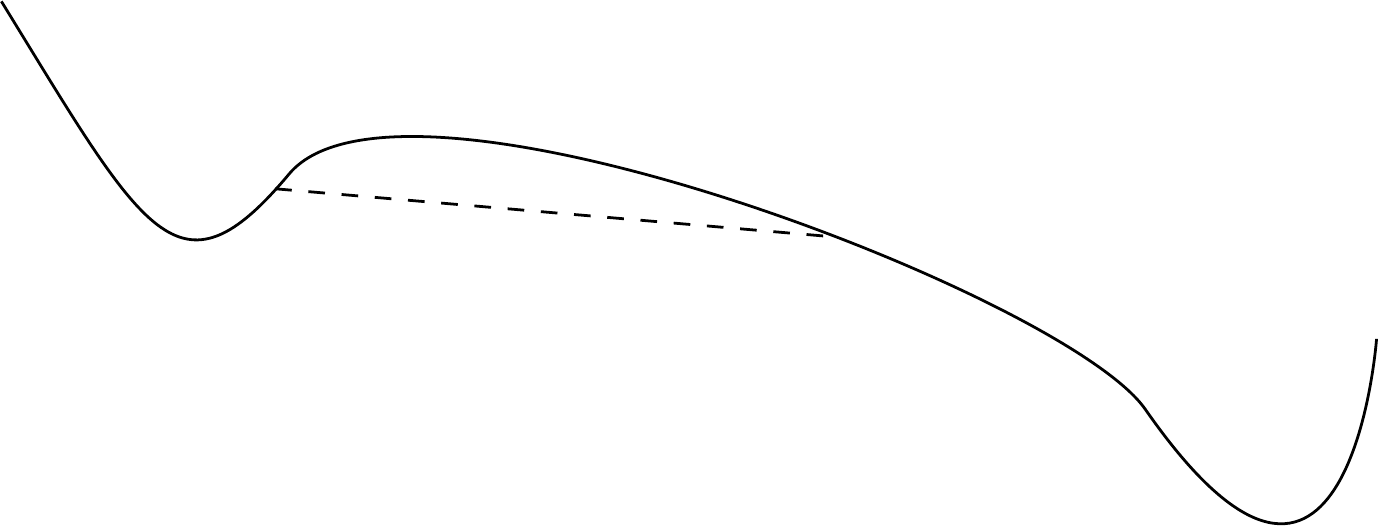}
\end{center}
\caption{A potential with a fat barrier.  The tunneling path is shown as a dashed line. For a broad enough peak the tunneling instanton will not exist, and the Hawking-Moss instanton will dominate, leading to slow-roll eternal inflation at the top of the barrier. We argue that potentials of this type are statistically suppressed in the landscape.}\label{fatbarrier}
\end{figure}

As a second application of the landscape framework, we will explain why we believe we are justified in taking the barrier to be quite sharp.\footnote{This discussion is inspired by conversations with Enrico Pajer.} By contrast, one could imagine a potential of the form shown in Fig.~\ref{fatbarrier}, in which the false vacuum is separated by a broad barrier. 
This construction however requires the top of the barrier to be quite flat.  More quantitatively we can argue that we must have $\eta\ll1$ at the top, by noting that as long as $V'<0$ we have
\begin{align}\nonumber
\frac{d}{d\phi}\log V&=-\frac{\sqrt{2\epsilon}}{M_P}\ , \\
\frac{d^2}{d\phi^2}\log V&=\frac{\eta-2\epsilon}{M_P}\  ,
\end{align}
and thus that
\be\label{etaint}
\sqrt{2\epsilon}=-\int_{\phi_{peak}}^\phi\frac{d\phi}{M_P}(\eta-2\epsilon).
\ee
We then observe that any $O(1)$ negative value of $\eta$ near $\phi_{peak}$ would need to be offset by an $O(1)$ positive value of $\eta$ at some later time over a comparable field range in order to get $\epsilon$ to be small again during the period of observable inflation.  This would lead to the kind of potential with steeping that we have considered so far. One could try to avoid this conclusion by asking instead that the field excursion from $\phi_{peak}$ to the beginning of inflation be small in Planck units, shrinking the range of the integration in \eqref{etaint} and thus allowing large values of $\eta$ near the top.  Doing this would require $\eta$ to get from $O(1)$ to a small value quite quickly and then stay there, which would need fairly precise manipulation of the higher derivatives of $V$. 
Once we have $|\eta|\ll1$ at the top of the hill, this requires additional statistical tuning beyond the one needed to get inflation. In general this ``wide-barrier'' scenario seems to require extra statistical tuning compared to the potential in Fig.~\ref{potentialfig}, with no added benefit. Incidentally $|\eta|\ll1$ at the top of the hill changes the physics qualitatively.  As reviewed for example in \cite{Batra:2006rz}, in this situation the Coleman-De Luccia instanton no longer exists and the relevant transition is now the Hawking-Moss instanton \cite{Hawking:1981fz}.  This instanton can be interpreted as causing an entire Hubble-sized region to thermally jump up to the top of the potential barrier \cite{Batra:2006rz,Brown:2007sd}.  Since $|\eta|\ll1$ and $\epsilon=0$ there, slow-roll eternal inflation then necessarily ensues. Thus the present issue is closely related to the previous subsection.

Of course the potential in Fig.~\ref{potentialfig} still has some non-generic features: the inflationary plateau is flat, its scale is low, and the cosmological constant in our vacuum is small.  Unlike the other statistical tunings just we've just discussed however, these can be anthropically explained. The anthropic argument for the smallness of $\rho_{\Lambda}$ is well known \cite{Linde:1984ir,Banks:1984cw,Weinberg:1987dv}.  The low scale of the inflationary plateau is needed to avoid over-production of structure, since we need $V_{\rm inflation}\approx 10^{-10}\epsilon$ in Planck units. The flatness of the plateau is needed to get enough e-foldings to dilute curvature \cite{Vilenkin:1996ar,Freivogel:2005vv}.  This last statistical tuning, the flatness of the potential, is quite close to the steepening we are interested in; it was studied quantitatively in \cite{Freivogel:2005vv}, and we now study the steepening along similar lines.

\section{The Probability for Observable Steepening}
\label{sec-probability}

In this section, we further quantify the notion that flat potential regions require statistical tuning. Our goal will be to estimate the probability for observing a power suppression in the CMB due to steepening of the potential near the beginning of inflation.  We will consider a range of plausible assumptions about the statistical distribution of potentials in the landscape. Again, as for the former section, it is worth stressing that only the conclusions of the present section depend on these assumptions.

In making statistical predictions from the landscape, it is important to clearly state which parameters are being varied and which are not.  In this section we assume that the power spectrum asymptotes at large $k$ to the standard $\Lambda$CDM power spectrum given by \eqref{Ppar}.  Motivated by our discussion of the previous sections, we will further assume that below some $k$ the power is suppressed.  At some even lower $k$, the potential has become steep enough that inflation is no longer possible; this is the beginning of inflation.  This model of the power spectrum involves essentially four parameters: the asymptotic value of $A_s$ and $n_s$, as well as the point of onset of power suppression and of the onset of inflation. Here we will take the asymptotic parameters $A_s$ and $n_s$ to be fixed.  We thus do not need to discuss prior distributions or anthropic cuts for these parameters. Of course a more general analysis would do so, but already in fixing them, there is still a chance that our discussion of the remaining two parameters might be in conflict with observations.\footnote{ We briefly comment however that it is unclear to what extent the prior for $A_s$ favors larger or smaller values.  On the one hand $V$ likes to be large, but on the other hand so does $\epsilon$, so the tendency of their ratio is unclear.  Making $A_s$ smaller prevents structure formation, so any tendency in this direction would be cutoff anthropically fairly quickly.  Making it larger is more subtle, but it has been argued that there is an anthropic cut in this direction as well \cite{Tegmark:1997in}.  Scanning the parameter $n_s$ is expected to be quite irrelevant. We leave a detailed discussion of this to future work.}   (If they are not, i.e., if we succeed for now, then later work may still falsify the theory by allowing $A_s$ and $n_s$ to vary and exhibiting a conflict between the predicted and observed values. In the present context, theory should be understood to include our assumptions about the prior distribution, Eqs.~\eqref{prior} below, which may require modification.)  Our project for the rest of this section is thus to motivate prior distributions for the location of the onset of potential steepening and the beginning of inflation, identify anthropic cuts on this two-parameter space, and then compute the probabilities for observable suppression or curvature.

\begin{figure}
\begin{center}
\includegraphics[height=6cm]{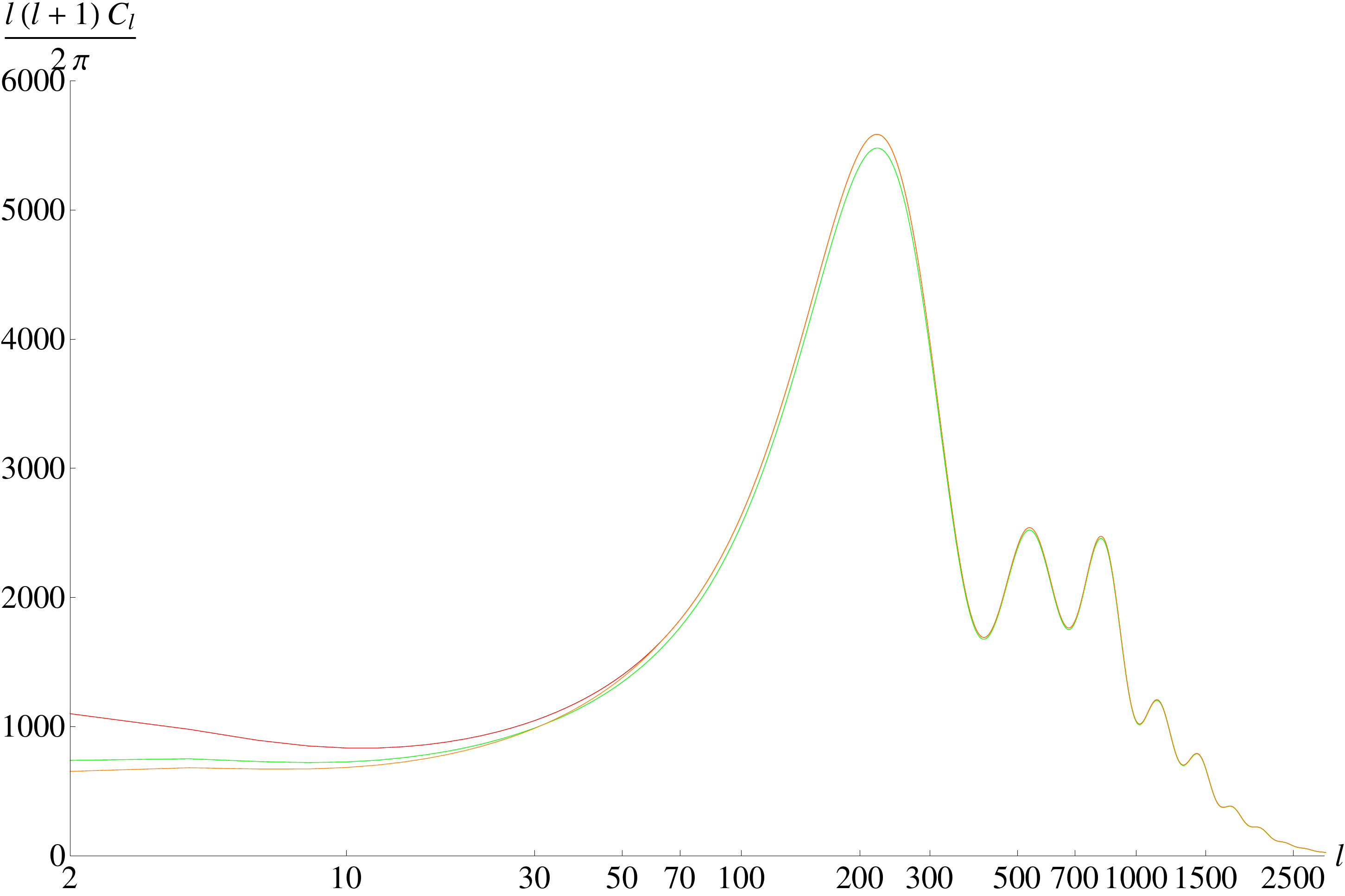}
\end{center}
\caption{The $TT$ power spectrum for the best fit to $\Lambda$CDM in red, our exponential model in green, and our power-law model in orange, all plotted on a log scale.}\label{logplot}
\end{figure}

A first worry is that, given that we have excluded steepening above $\ell\approx 100$, the remaining window is small enough that seeing steepening is very unlikely.  In looking for features of the inflationary potential in the CMB, a crucial point is that the relationship between $\phi$ and $\ell$ is logarithmic, as can be seen from equation \eqref{phik} and the discussion below it.  We emphasize this in Fig.~\ref{logplot}, where we plot the power spectrum for $\Lambda$CDM and for the two models of Sec.~\ref{models} against $\log \ell$.  
For example, the $C_\ell$'s with $2<\ell<70$ contain information about a region of the inflationary potential which is just as large as the region described by the $C_\ell$'s with $70<\ell<2500$. We will argue below that the probability distribution over the location of the steepening feature, in terms of $\log\ell_s$, is not very steep in this regime. Moreover, we will argue that the entire observable regime is anthropically allowed. Thus, we will find that the probabilities for finding a steepening feature, say, in the range $2\lesssim \ell< 70$ is not much smaller than finding it in the range $70<\ell_s<2500$.

Another question is why a steepening feature should lie in the region $2<\ell<2500$ in the first place. Why should it be located where it can be observed through the CMB? The angular modes that leave the horizon during inflation range up to exponentially large values ($\ell_{reh}\approx e^{63}$ for GUT-scale inflation). Moreover, if there were extra e-folds of inflation beyond what is needed to dilute curvature, then there would also be room at ``$\ell\ll1$'' for the steepening to happen. This regime is invisible because the corresponding modes are too far outside our horizon. In Sec.~\ref{sec-ab}, we will describe an anthropic boundary which prevents the steepening feature from turning on at $\ell \gg 10^{4}$.  

In Sec.~\ref{sec-prior}, we will discuss prior distributions for the number of e-folds of inflation, and for the onset of steepening. They quantify our assumption that having the suppression turn on in the invisible ``$\ell\ll1$'' region is statistically suppressed, since it would require a larger total number of e-foldings of inflation.   

Combining these assumptions and results in Sec.~\ref{sec-cond}, we conclude that the probability for a steepening feature lying in the observable region can be as much as ${O}(1)$, while the probability for a curvature lying in the observable region is of order ${O}(10\%)$. This probability is larger than the probability for observable curvature, because the relevant anthropic boundaries correspond to different values of $\ell$; and it remains larger after observational exclusions are taken into account.

\subsection{Anthropic Bound On Steepening}
\label{sec-ab}

If flat potentials are rare, as we assume, the steepening should begin at very large~$\ell$. But there could be an anthropic reason why this cannot be the case.  The power spectrum of scalar density perturbations generated during inflation is proportional to $H^4/\dot{\phi}^2$ evaluated at the time where the mode of interest is leaving the horizon. A steeper potential will lead to a larger $\dot{\phi}$ and thus to a lower power spectrum~\cite{Contaldi:2003zv,Freivogel:2005vv,Yamauchi:2011qq}.  
If this happens at distance scales which are associated with the mass of a typical galaxy, it will prevent the formation of galaxies.

To see this quantitatively we need to briefly recall how primordial density perturbations are connected to structure formation.  In evolving modes from horizon entry to last scattering, it is important to know whether or not they were inside the horizon at the time of matter-radiation equality.  To understand which modes we are interested in, we note that the proper distance to the last-scattering surface today, equation \eqref{Dls}, is about 14 gigaparsecs.  The typical distance between galaxies is of order a few megaparsecs \cite{Weinberg:2008zzc}.  From equation \eqref{approxl} the relationship between the proper wavelength $\lambda$ of a fluctuation today and the angular scale of that fluctuation on the CMB is
\be
\ell\approx\frac{2\pi D_{ls}}{\lambda},
\ee
so the density perturbations important for the formation of galaxies evolved from fluctuations whose angular size on the CMB is of order 
\be
\ell_g\approx 10^{4.5}.
\ee
By comparison modes which were entering the horizon at matter-radiation equality have
\be
\ell_{EQ}=H_0 D_{ls}\frac{a_{EQ}H_{EQ}}{a_0H_0}\approx 143,
\ee
so the modes of interest for galaxy formation entered the horizon long before matter-radiation equality.  The cold dark matter density perturbation for such modes at the time of last scattering is given by\footnote{This equation follows from a hydrodynamic treatment of the evolution of density perturbations from horizon entry to last scattering; see equations 6.5.11 and 6.5.15 of \cite{Weinberg:2008zzc}. The $(\ell_{EQ}/\ell_{LS})^2$ factor comes from the linear (in the scale factor) growth of the perturbation after matter domination. The log comes from the growth during radiation domination.}
\be
\frac{\delta\rho}{\rho}\Big|_{t=t_{ls},k=\ell/D_{ls}}\approx\frac{9}{2}\left(\frac{\ell_{EQ}}{\ell_{LS}}\right)^2\log\left(\frac{4}{\sqrt{3}}\frac{\ell}{\ell_{EQ}}\right)\sqrt{P_s(\ell /D_{ls})}.
\ee
Here $\ell_{LS}\approx \sqrt{(1+z_{LS})\Omega_M} H_0 D_{ls}\approx 59$ is the angular scale of modes which were entering the horizon at the time of last scattering, and $P_s(k)$ is the primordial power spectrum.  This is the density perturbation for an individual mode, to get the typical size of the total mass fluctuation in a box of linear size $\frac{2 \pi}{k}$ we should sum in quadrature over the modes in the box whose wavelengths are of order $\frac{2\pi}{k}$.  This amounts to multiplying by a factor of $k^{3/2}$; for $\ell$ of order $\ell_g$, this gives a total mass perturbation of order $10^{-3}$ since $\sqrt{k^3 P(k)}$ is of order $10^{-5}$.

During the matter dominated era these perturbations grow like the scale factor $a\propto (1+z)^{-1}$,\footnote{See for example section 8.2 of \cite{Weinberg:2008zzc}.} so since $z_{ls}$ is of order $10^3$, $\frac{\delta\rho}{\rho}$ would become order one and the overdensity would become gravitationally bound at redshift $z\approx 1$. But this is when the cosmological constant began dominating the evolution of the universe and turning off perturbation growth. 

We then can see that galaxy formation in our universe is occurring close to a catastrophic boundary. Suppose that the initial density contrast was smaller, even by just a factor of order one. Then only exponentially rare multi-sigma overdensities would have formed large galaxies, and there would be exponentially fewer observers (according to all viable measures currently under consideration~\cite{BouMai12}). Smaller galaxies are further from the catastrophic boundary, but cooling problems impose a lower bound on galaxy mass~\cite{BouHal09}, and metallicity may be insufficient if hierarchical structure formation is disrupted too early.

The above argument makes the simplifying assumption that the change of slope, at the onset of steepening, is immediately so large as to suppress the density contrast by a factor of order one.  A more natural expectation is that the steepening sets on more gradually, ramping up from undetectable, to detectable, to order-one suppression of the power as $\ell$ decreases. It is difficult to parametrize such features in detail. But it is clear that this effect moves the anthropic boundary for the onset of detectable steepening towards even larger values of $\ell$. We will neglect this shift in what follows. This makes our estimates below for the probability of detecting steepening more conservative, but a larger uncertainty arises in any case from our ignorance of the details of the prior distribution.

\subsection{Prior Distribution for Steepening and Curvature}
\label{sec-prior}

From an observational viewpoint, the steepening feature is conveniently parametrized by the angular scale $\ell_s$ at which it appears in the CMB spectrum. However, from a theoretical perspective, the feature appears in the slow-roll potential. Thus, its prior distribution is more naturally parametrized in terms of the number of e-folds before the end of inflation, $\Ss$, at which the potential steepens noticably. This number is related simply to the potential via equation \eqref{phik}.

The logarithmic relationship \eqref{phik} between $k$ and $\phi$ implies
\be
\Ss\equiv \log{\frac{\ell_{reh}}{\ell_s}}\approx 63-\log \ell_s~.
\label{eq-convert}
\ee
Here the index $reh$ indicates reheating, and in the second equality we have assumed reheating at the GUT scale. The anthropic cutoff of the previous section demands $\Ss>\Ss_g\equiv 52$.  

Following inflation back in time, steepening indicates the transition between a flat slow-roll potential and a steep barrier separating our vacuum from its parent vacuum. Slow-roll inflation can begin at a somewhat more negative value of $\phi$ than the location of this feature, but one expects that the location of the steepening feature is statistically distributed in a way very similar to the distribution of the total number of e-folds of inflation, $\mathcal{N}$.  For motivation we now briefly recall this distribution.

Based on our general discussion of statistical tuning, the prior probability distribution for $\mathcal{N}$ in the landscape should be suppressed at large $\mathcal{N}$. A plausible guess is that at large $\mathcal{N}$ the prior distribution should fall like some power 
\be
\frac{dP}{d\mathcal{N}}\propto \mathcal{N}^{-\nu},
\label{eq-nn}
\ee
with $\nu>1$ but of order unity \cite{Freivogel:2005vv}. (Indeed for a particular parametrization of a linear potential, assuming a uniform distribution over parameters implies $\nu=4$ \cite{Freivogel:2005vv}, while $\nu\approx 3$ in a mini-landscape inspired by a brane moving in a conifold~\cite{Agarwal:2011wm}.) Exponential suppression is ruled out experimentally, since slow-roll inflation would be too rare in the landscape to explain the observed flatness of the universe even after anthropic conditioning.

A natural first guess for the prior distribution for the steepening location $\Ss$ is that at large $\Ss$ it falls like $\Ss^{-\zeta}$ for some $\zeta>1$ of order unity, but this does not quite work since in fact $\mathcal{N}$ and $\Ss$  are not independent; the beginning of inflation must happen at more negative $\phi$ than the onset of steepening.  We can implement this as a constraint $\mathcal{N}\geq \Ss$, but this means that in fact we need to consider a joint distribution for both $\Nn$ and $\Ss$.  We thus propose an unnormalized joint distribution\footnote{We have also considered some other similar distributions, the results are comparable to what we present here.}
\be\label{prior}
\frac{dP}{d\Nn d\Ss}=\Ss^{-(\zeta+1)}\left(\frac{\Ss}{\Nn}\right)^\nu \Theta(\Nn-\Ss)\ .
\ee
with $\nu$ and $\zeta$ being order one positive numbers.
The first term in this distribution suppresses large values of $\Ss$, as expected from our general discussion of flatness, while the second implements the expectation that once the potential begins to steepen, it typically should not be too much longer before it can no longer support inflation.  The parameterization is chosen so that integrating out $\Nn$, without imposing anthropic constraints, gives $\frac{dP}{d\Ss}\propto\Ss^{-\zeta}$; and integrating out $\Ss$, again without anthropic constraints, gives $\frac{dP}{d\Nn}\propto \Nn^{-\gamma}$, with $\gamma=\zeta-\nu$, so the suppression for $\Nn$ is smaller than for $\Ss$.  Larger $\zeta$ suppresses large $\Ss$ more, making observable power suppression more likely. $\nu$ governs the expected separation between power suppression and curvature.  If $\nu$ is large, then potentials like our exponential model should be more typical. If $\nu$ is small, then our power law model should be more typical.

\subsection{Conditional Probability Distributions}
\label{sec-cond}

We now turn to the computation of the conditional probability distributions for the location of the steepening feature, $\Ss$, and the number of e-foldings, $\Nn$, after conditioning on the existence of galaxies, and later, in addition, on observational exclusion limits. This will allow us to estimate the probability for a discovery of either feature in the CMB. In this analysis we will need the corresponding anthropic boundary and exclusion limit for curvature, which we now briefly review.  

The anthropic constraint that curvature does not disrupt galaxy formation requires $a_{LS}H_{LS}>10$~\cite{Freivogel:2005vv}.  Assuming GUT-scale inflation,\footnote{The appearence of the GUT scale here has nothing to do with particle physics.  The scale is really chosen by taking the highest possible scale for inflation that has $P(k)$ of order $10^{-10}$ and $\epsilon$ at most $10^{-2}$.  Using the quoted parameters below equation \eqref{Ppar}, this gives reheating energy density $\rho\approx (2\times 10^{16} GeV)^4$. It is straightforward to check that making the scale of reheating high is the ``worst-case'' assumption from the point of view the probabilities for observing steepening and curvature we compute below, but the dependence is logarithmic so the scale has to be decreased a lot to improve the odds significantly.} this condition becomes
\be
\mathcal{N}>61\; .
\ee
The observational constraint is of course stronger. The number of e-foldings required to dilute curvature to the current exclusion limit, $\Omega_K<10^{-2}$, is $\mathcal{N}\approx 64$.  Curvature becomes unobservable in principle \cite{Kleban:2012ph} when $\Omega_K\lesssim 10^{-4}$, which happens for $\mathcal{N}\gtrsim 66$.

Note that the window in which the steepening is both anthropically allowed and potentially observable, $52<\mathcal{S}<63$, is about twice as  wide as the analogous window for curvature, $61<\mathcal{N}<66$.  (And the difference would be even greater, had we not assumed the worst case, that the change in slope is so dramatic as to completely eliminate galaxy formation, for a steepening feature located at $\Ss_g$.) As a result, we are about to find that the probability for observing steepening is about twice as large as the probability for observing curvature.  Why is the window for steepening larger, even though both anthropic boundaries are related to galaxy formation?

The reason is that curvature, like a positive cosmological constant, disrupts the {\em growth} of small perturbations. It can do so at any time while perturbations are still in the linear regime, nearly until virialization. Thus, if curvature dominates the evolution of the scale factor before perturbations on the galactic scale become nonlinear, galaxies will not form. Steepening, on the other hand, only affects the {\em initial strength} of the perturbations. Consequently, the anthropic bound from curvature is set by the horizon scale at virialization, which corresponds in the CMB to $\ell\sim O(1)$, whereas the bound associated with steepening is set by the much smaller comoving scale of the current intergalactic distance, corresponding to $\ell\sim \ell_g=10^{4.5}$.  

Treating both anthropic cutoffs as sharp, the probability distributions conditioned on the existence of galaxies can be obtained by removing the forbidden region and renormalizing.  We can then compute the probabilities that the power suppression or curvature lie in their respective observable windows $52<\Ss<63$ and $61<\Nn<66$:
\begin{align}\nonumber
P_{\rm suppression}&=\frac{\int_{52}^{63} d\Ss \int_{61}^\infty d\Nn \frac{dP}{d\Ss d\Nn}}{\int_{52}^\infty d\Ss \int_{61}^\infty d\Nn \frac{dP}{d\Ss d\Nn}}\\
P_{\rm curvature}&=\frac{\int_{52}^\infty d\Ss \int_{61}^{66} d\Nn \frac{dP}{d\Ss d\Nn}}{\int_{52}^\infty d\Ss \int_{61}^\infty d\Nn \frac{dP}{d\Ss d\Nn}},\label{condprobs} 
\end{align}
where $\frac{dP}{d\Ss d\Nn}$ is the prior \eqref{prior}.  (The probability for seeing both curvature and power suppression is nearly as large as the probability for seeing curvature.)  

\begin{figure}
\begin{centering}
\includegraphics[height=4cm]{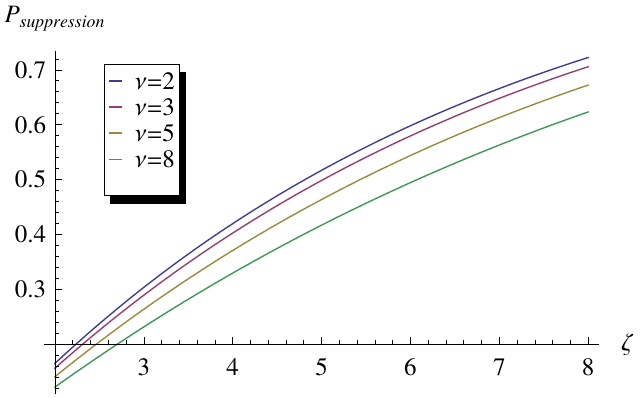}\hspace{.7cm}\includegraphics[height=4cm]{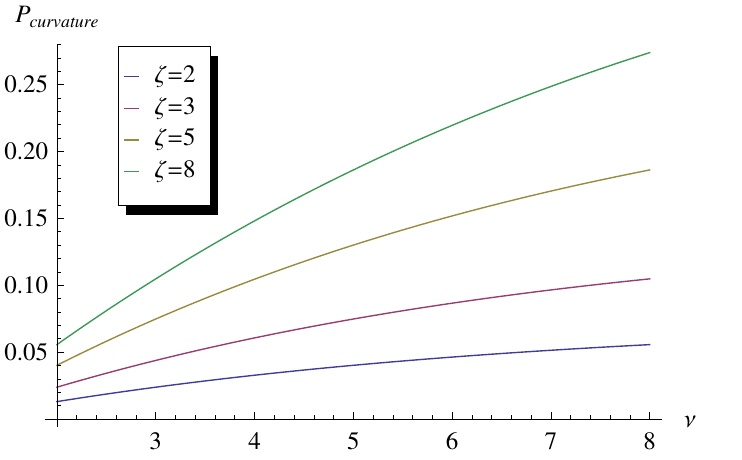}
\caption{On the left, a plot of $P_{\rm suppression}$, the probability for observable power suppression, as a function of $\zeta$, for various values of $\nu$.  On the right, a plot of $P_{\rm curvature}$, the probability for observable curvature, as a function of $\nu$, for various values of $\zeta$. These are total probabilities that do not take into account that some signal regions have already been excluded.}\label{probs}
\end{centering}
\end{figure}

$P_{\rm suppression}$ and $P_{\rm curvature}$ are plotted for some ranges of $\nu$ and $\zeta$ in figure \ref{probs}.
As expected, the probability for seeing power suppression is significantly larger than the probability for seeing curvature.  For $\nu=3$ and $\zeta=4$ it is about $40\%$, and it can be even larger.  It may be somewhat surprising that increasing $\nu$ at fixed $\zeta$, which makes seeing curvature more likely, makes seeing power suppression less likely.  This is because the term $\left(\frac{\Ss}{\Nn}\right)^\nu$ in the prior distribution \eqref{prior} suppresses a large separation between $\Nn$ and $\Ss$, which pushes $\Ss$ up towards the anthropic cut for $\Nn$ once we impose it, even though the cut for $\Ss$ is significantly lower.

\begin{figure}
\begin{centering}
\includegraphics[height=4cm]{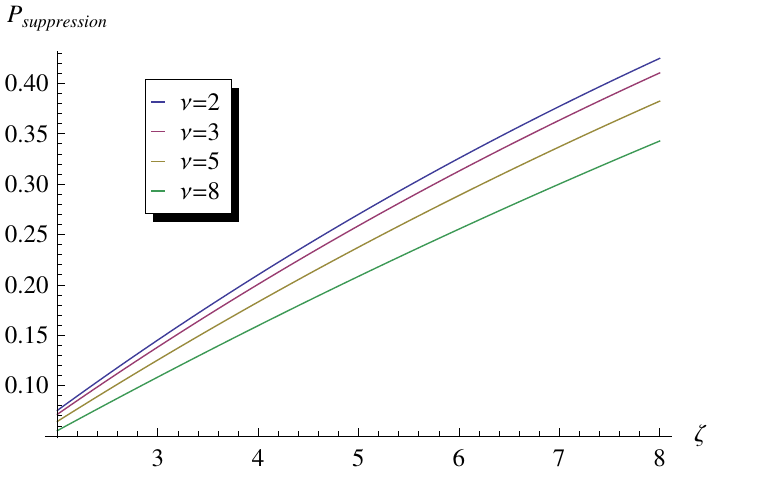}\hspace{.7cm}\includegraphics[height=4cm]{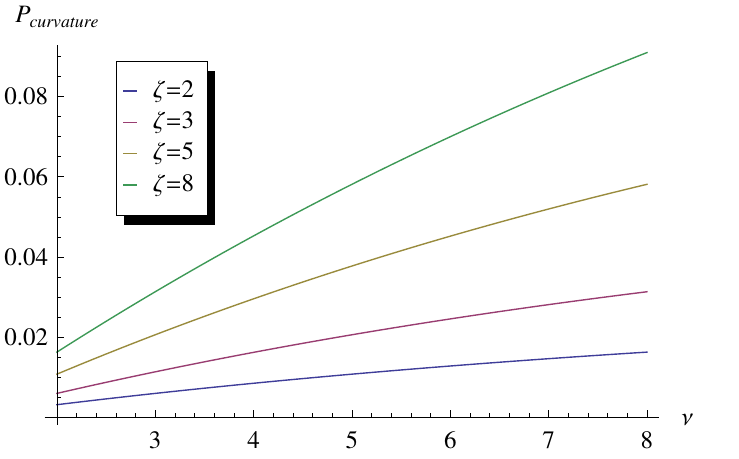}
\caption{Plots of $P'_{\rm suppression}$ and $P'_{\rm curvature}$, obtained after conditioning on current observational exclusions. These are the probabilities that future experiments will find a signal.
}\label{probscon}
\end{centering}
\end{figure}

We can also compute the probabilities $P'$ for future detection of power suppression or curvature, given the current exclusions.  $\mathcal{N}\gtrsim64$ is required for $\Omega_K\lesssim 10^{-2}$, while $\Ss \gtrsim 58$ is needed for the suppression to begin below $\ell\approx 100$ as suggested in section \ref{sec-exclude}.  We can include these constraints simply by replacing $61\to64$ and $52\to58$ in \eqref{condprobs}.  We show the results in figure \ref{probscon}; as expected they both decrease by about a factor of two.

These estimates are quite rough. The odds for a future discovery improve if we take into account that the steepening feature may turn on gently at $\ell\gg \ell_g$ and slowly ramp up. This shifts the anthropic boundary to values below $\Ss_g=52$. On the other hand, the odds for an observable feature decrease if the steepening is so subtle that it escapes detection. Moreover, cosmic variance will limit our ability to detect a feature close to $\ell\sim 1$ ($\Ss\approx 63$), so it may be appropriate to narrow the range of $\Ss$ that can be considered observable. One could also weaken the anthropic bounds somewhat by allowing observers to form in smaller galaxies, but this does not significantly change our results.   $S$ depends only logarithmically on scale, so decreasing the galaxy scale even by a factor of 10 gives only a small (3-5\%) increase in the probability of seeing suppression.  Moreover this is only true for the probability {\it before} imposing observational constraints. Since the observational constraint is already at lower $\log \ell$ than the anthropic bound, relaxing the latter to larger $\log \ell$ has no effect on the probability for future detection.

It is worth emphasizing an important point again. Because the probability distribution for the steepening location is roughly flat not over $\ell$ but over $\log\ell$, we should {\em not\/} necessarily expect the steepening feature to be close to the anthropic boundary, in the sense of $\ell\sim 10^{4.5}$.  For example, assuming an observable steepening feature and $\zeta=4$, the probability that the feature lies at $\ell<50$ is about $70\%$.

\acknowledgments We would like to thank Julien Lesgourgues for help understanding and using his CLASS CMB Boltzman code, and especially for modifying it to include one of our models. DH would like to thank Xiaowei Yu for help in understanding hypothesis testing.  We thank Adam Brown and Alex Dahlen for very helpful comments on CDL instantons, and Matias Zaldarriaga for pointing out a mistake in an earlier version of section 3.2.  We are grateful to Sergei Dubovsky, Daniel Green, Ben Freivogel, Thomas Hertog, Matt Kleban, Andrei Linde, Samuel Lee, Juan Maldacena, Daan Meerburg, Enrico Pajer, Michael Salem, Uros Seljak, Steve Shenker, Eva Silverstein, David Spergel, Paul Steinhardt, Lenny Susskind, Timm Wrase, and Matias Zaldarriaga for helpful discussions. DH and RB would like to thank CERN for hospitality during a workshop when this work was initiated, and DH and LS would like to thank KITP for hospitality and a stimulating environment at the ``Primordial Cosmology'' workshop.  DH and RB would also like to thank the KITP for hospitality during the Fuzz or Fire workshop.  This work was supported by the Berkeley Center for Theoretical Physics, by the National Science Foundation (award numbers 1002399, 0855653 and 0756174), by fqxi grant RFP3-1004, by ``New Frontiers in Astronomy and Cosmology'', by the Princeton Center for Theoretical Science, and by the U.S.\ Department of Energy under Contract DE-AC02-05CH11231.  Leonardo Senatore is supported by DOE Early Career Award DE-FG02-12ER41854 and by NSF grant PHY-1068380.

\appendix

\section{Statistics for String Theorists}\label{statistics}
There is a simple way to get a rough estimate for the significance of the suppression we have been discussing. The basic idea is to calculate the likelihood of the data assuming that the measured $C_\ell$'s are independently Gaussian distributed about the red curve in Fig.~\ref{datafig2}.  The $\chi^2$ variable
\be
\chi^2\equiv \sum_{\ell=2}^{\ell_{max}}\left(\frac{C_{\ell}^{\{data\}}-C_{\ell}^{\{theory\}}}{\sigma_{\ell}}\right)^2,
\ee
where we take the standard deviations $\sigma_\ell$ of the Gaussians to be the average of the distances of the upper and lower error bars in Fig.~\ref{datafig2} from the observed value, gives a decent measure of the absolute deviation of the data away from the red theory curve.\footnote{A subtlety which was emphasized to us by Uros Seljak is that most sources of error are proportional to the signal.  So in comparing models other than $\Lambda$CDM to the data we should rescale the standard deviations $\sigma_\ell$ by a ratio of the $C_\ell$ for the model over $C_\ell$ for $\Lambda$CDM with the best-fit parameters.  We take this into account in the significances reported in section \ref{models}, although the effect turns out to not be too large.}    $\ell_{max}$ is a parameter which controls over what range we are looking for an effect.  With our Gaussian probability assumption we can compute the $p$-value for a deviation at least this large from the cumulative $\chi^2$ distribution:\footnote{See for example the review of probability and statistics from \cite{Beringer:1900zz}, although it is straightforward to derive this from the Gaussian distribution by integrating out the angular directions.}
\be
p_{\chi^2}(\chi^2)\equiv \frac{1}{2^{\frac{\ell_{max}-1}{2}}\Gamma\left(\frac{\ell_{max}-1}{2}\right)}\int_{\chi^2}^\infty dx\, x^{\frac{\ell_{max}-3}{2}}e^{-x/2}.
\ee
In looking for a suppression we must also take into account the sign of the deviations, which the $\chi^2$ parameter is insensitive to.  For our Gaussian distribution this sign is uncorrelated with the absolute size of the fluctuation, so we can take the probability that at most $m$ points lie above the red curve to be given by a binomial distribution
\be
p_{\pm}(m)=\frac{1}{2^{\#(S)}}\sum_{k\in S,k\leq m} \begin{pmatrix}
\#(S)\\
k
\end{pmatrix},
\ee
where to make this estimate more robust we have included only $\ell$'s in a set $S$ containing those $\ell<\ell_{max}$ whose $C_{\ell}$'s differ from the theory curve nontrivially.\footnote{This may seem arbitrary, but in any event all of the six excluded $\ell$'s below 30 and five out of the seven excluded between 30 and 50 lie below the curve, so including them would only increase the significance of the suppression.}
The full $p$-value for estimating the significance of the suppression is then $p\equiv p_{\chi^2}(\chi^2)p_{\pm}(m)$, from which we can determine the number of standard deviations by solving
\be
p=\frac{1}{\sqrt{2\pi}}\int_n^\infty e^{-x^2/2}.
\ee
Applying this method to the model and data in Fig.~\ref{datafig2} we find $n=1.8$ for $\ell_{max}=49$ and $n=2.4$ for $\ell_{max}=30$, which is roughly consistent with the much more sophisticated analysis of \cite{Planck:2013kta}.  We emphasize that this method is no substitute for a real Monte Carlo analysis scanning over parameters, although we believe it suffices for the theoretical points we make here.

\section{Where does the CDL Instanton Drop Us?}\label{dropapp}
\begin{figure}
\begin{centering}
\includegraphics[height=5cm]{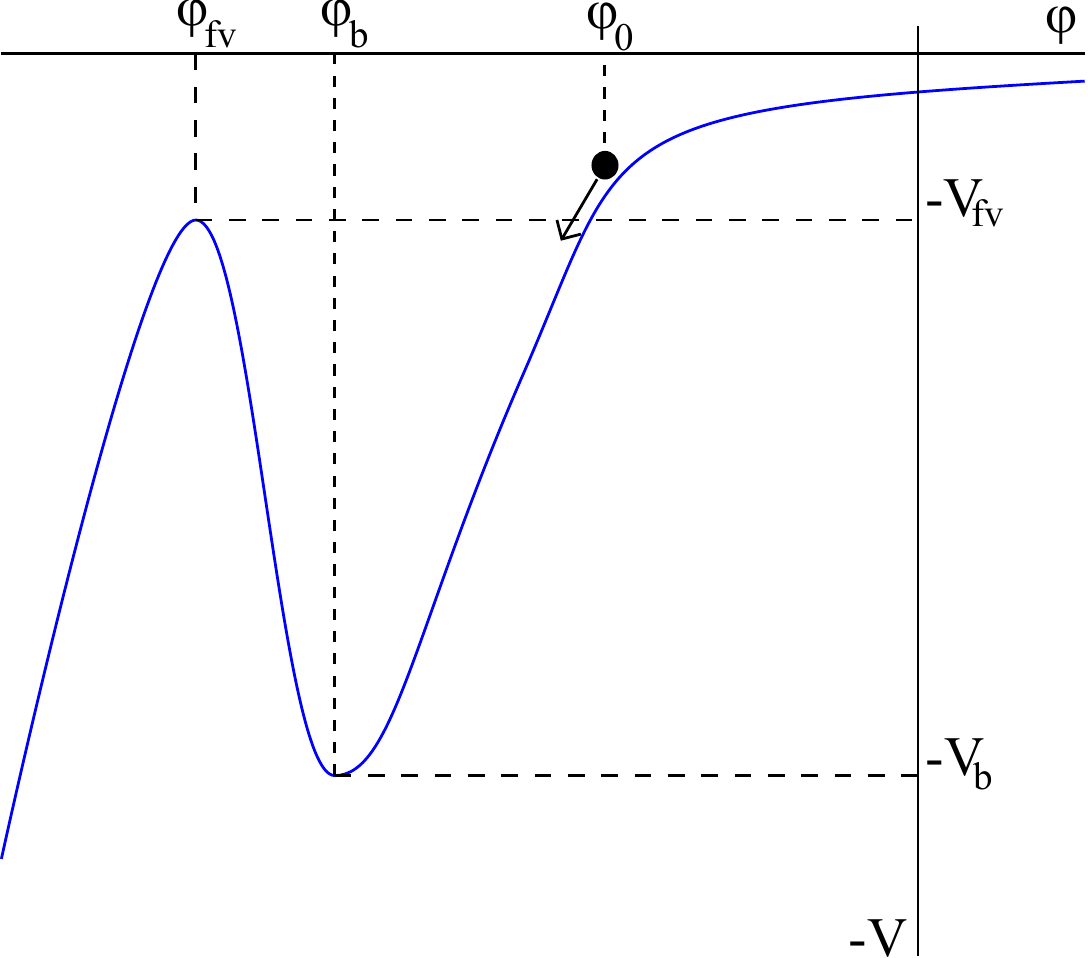}
\caption{Evolution on the potential $-V$.  To overcome friction we must start the field higher than $-V_{\rm fv}$, after which it rolls down through the barrier at $\phi_b$ and comes up back up to rest at $\phi_{\rm fv}$ at Euclidean time $\xi=\infty$.}\label{minuspot}
\end{centering}
\end{figure}

In this appendix we argue that for potentials like that of Fig.~\ref{potentialfig}, the tunneling instanton typically drops the field off high enough that the subsequent Lorentzian evolution rolls through enough of the steepening region to produce a potentially observable power suppression.  For simplicity we first consider the $M_p\to \infty$ limit, where the geometry becomes Minkowski space and the Euclidean instanton is a solution of the equation
\be\label{bounceeq}
\phi''+\frac{3}{\xi}\phi'=V',
\ee
where $V$ is some potential of the form shown in Fig.~2 and $\xi$ is Euclidean time.  The boundary conditions are that $\phi'(0)=-1$ and $\phi(\infty)=\phi_{\rm fv}$.  For intuition we can reinterpret $\xi$ as a Lorentzian time, in which case this equation describes the motion of a particle with potential $-V$, with a friction term.  This situation is illustrated in Fig.~\ref{minuspot}.  The instanton is a trajectory which starts at rest at $\xi=0$ and $\phi=\phi_0$, rolls ``down'' through the potential barrier, and then climbs back ``up'' to come to rest on top of the false vacuum at $\xi=\infty$.  If the friction term were absent then by energy conservation it is obvious that we would need to choose $\phi_0$ such that $V\left(\phi_0\right)=V_{\rm fv}$.  Since $\phi_0$ is also the starting point for real Lorentzian evolution inside the bubble, and since we expect that $V_{\rm fv}\gg V_{\rm inf}$, where $V_{\rm inf}$ is the scale of inflation, this would ensure that the Lorentzian evolution of the field begins at a scale much higher than the inflationary plateau, leaving plenty of room to roll through the steepening feature on the way to the inflationary plateau.  Because of friction however, we need to start the field higher up on \textit{minus} the potential, which means that the tunneling event drops the field \textit{lower} on the potential.  

One may worry that because of friction it is actually necessary to have $\phi_0$ already on the inflationary plateau.  If this were generically what happened, then the steepening of the potential would not lead to an observable effect in the CMB since the field would not roll through it inside the bubble.  It may appear ``obvious'' from Fig.~\ref{minuspot} that, since $V_{\rm fv}\gg V_{\rm inf}$, friction should not be important enough to move the starting point all the way from $V_{\rm fv}$ to something of order $V_{\rm inf}$.  Adam Brown and Alex Dahlen have emphasized however that this conclusion depends on the form of the potential barrier.  In particular if the height of the barrier $V_b$ is parametrically larger than $V_{\rm fv}-V_{\rm inf}$, then from the point of view of solving equation\eqref{bounceeq} we should think of the the false vacuum and the inflationary plateau as essentially being degenerate, even if $V_{\rm inf}/V_{\rm fv}\ll1$.  From Fig.~\ref{minuspot} it would then seem quite likely that friction would require the field to be dropped off on the inflationary plateau.  This limit seems statistically tuned to us, but still this argument suggests that if we take $V_b\approx V_{\rm fv}-V_{\rm inf}\approx V_{\rm fv}$ then we should be right at the edge of this limit and it should be ambiguous whether or not $\phi_0$ is on the inflationary plateau.  

In fact the situation is better than it appears.  For anthropic reasons the potential barrier is quite asymmetric; on the left side is a high-scale false vacuum whose parameters should typically be Planckian, while on the right side it interpolates to a very flat region in order to get inflation and structure formation.  As long as this interpolation does not happen extremely sharply, we expect that the field will be dropped off high enough that it will be able to roll through the last part of the interpolation and thus produce a potentially observable suppression of power.  We now briefly present two models that support this intuition.

\begin{figure}
\begin{centering}
\includegraphics[height=5cm]{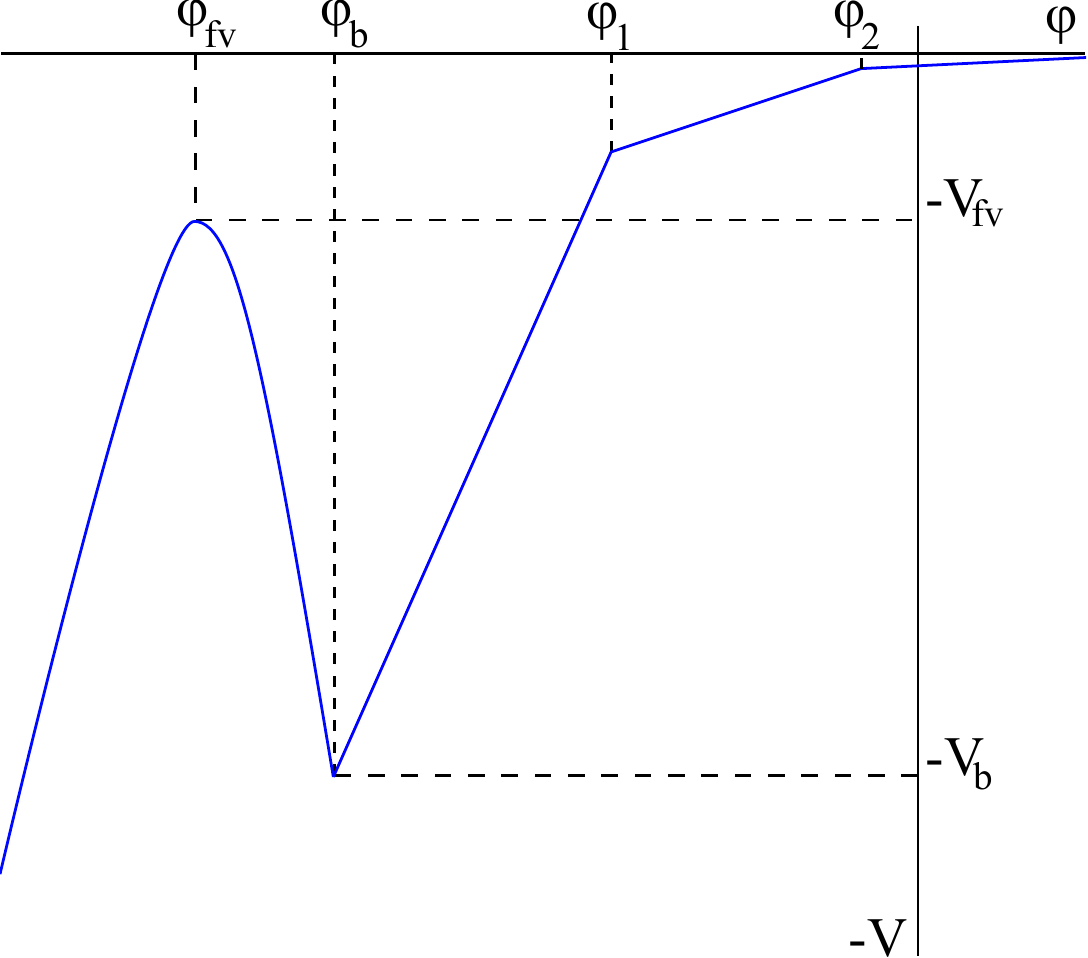}
\caption{A piecewise potential.  For $\phi<\phi_b$ it is quadratic, and there are three linear segments of decreasing steepeness as $\phi$ increases.}\label{linearquadpot}
\end{centering}
\end{figure}

For the first model, we consider the piecewise linear/quadratic potential shown in Fig.~\ref{linearquadpot}.  For this potential the instanton can be constructed analytically by sewing together solutions of \eqref{bounceeq} in the different regions.  We won't describe this explicitly, but the main results are as follows.  First of all if $V_b>4V_f$ then we must have $\phi_0>\phi_1$.  So if we had tried to only have two linear segments, meaning that the interpolation was instantaneous, this would realize the Brown-Dahlen claim that a reasonably high barrier requires the instanton to drop us on the flat region.  If we insert an extra segment, which here we take be in the region $\phi_1<\phi<\phi_2$, the situation is more interesting.  In particular in the regime where $V_b\gg V(\phi_1)\gg V(\phi_2)$, with $\phi_b-\phi_{\rm fv}\ll \phi_1-\phi_b$, we find that as long as $\phi_2-\phi_1$ is at least of order $\phi_1-\phi_b$ then $\phi_0$ is greater than but close to $\phi_1$.  The Lorentzian evolution then rolls through most of the region $\phi_1<\phi<\phi_2$, giving potentially observable steepening.  This is perhaps the most natural region of parameter space, since it represents a more smooth interpolation between the the steep barrier and the flat plateau.  

Of course one might worry that this argument requires to interpolation to be so gradual that the steepening would not be observable.  In our second model we show that this is not the case.  For our second model, for $\phi>\phi_b$ we take the potential to be 
\be
V_S+\gamma V_R,
\ee
with $V_S$, $\gamma$, and $V_R$ taken from the exponential model we discuss in section \ref{sec-exp}, with the parameters the same as were used to fit the \textsl{Planck} anomaly.  For $\phi<\phi_b$ we take the potential to be quadratic
\be
V=\frac{1}{2}m^2(\phi-\phi_{\rm fv})^2+V_{\rm fv},
\ee
with $\phi_{\rm fv}$ chosen to make the potential continuous at $\phi_b$.  In this model inflation begins at $\phi\approx 0$, which is also where we took horizon crossing to be in section \ref{sec-exp}, so as long as we find $\phi_0$ less than zero the field will roll through the steepening region that led to CMB power suppression.  We will take $V(\phi_b)=10^{-4}M_p^4$, $m=10^{-1}M_p$, and then study $\phi_0$ as a function of $V_{\rm fv}$.  Solving equation \eqref{bounceeq} numerically, we find that if we take $V_{\rm fv}=10^{-1}V_b$ then the field is dropped at $\phi_0/M=-9.6$, while if we take $V_{\rm fv}=10^{-2}V_b$ it is dropped at $\phi_0/M=-3.6$.  In both cases $\phi_b/M\approx16$ and $\phi_{\rm fv}/M\approx 17$, so indeed $\phi_b-\phi_{\rm fv}\ll \phi_{b}$ and the barrier is quite asymmetric.  Thus we seem to be ok even in the somewhat unnatural case where $V_{\rm fv}/V_{b}\approx 10^{-2}$.  Since the fairly steep case of an exponentially growing potential already works, we should also be fine for our more gentle power-law steepening model from section \ref{sec-inverse}.

Finally let's consider including dynamical gravity.  The only change to the $\phi$ equation of motion is to change $\frac{1}{\xi}$ to $\frac{a'}{a}$, with $a$ determined by simultaneously solving the Euclidean FRW equation
\be
\left(\frac{a'}{a}\right)^2=\frac{1}{a^2}+\frac{1}{3M_p^2}\left(\frac{1}{2}\phi'^2-V\right).
\ee
Instead of the instanton coming to rest at the false vacuum at $\xi=\infty$, it now comes to rest earlier at a point $\xi_c$ where $a$ linearly returns to zero.  It is not hard to see that the friction is now less than in the gravity-free case we just studied. Indeed observe that the Euclidean Hubble constant $H=\frac{a'}{a}$ obeys $H'=-\left(\frac{1}{a^2}+4\pi G \phi'^2\right)$, so the positivity of $\phi'^2$ ensures that $H$ is always less than it would have been with $G=0$.  This means that as we slowly turn on gravity, the drop-off point $\phi_0$ will have to move \textit{up} the potential (or down minus the potential).  This only makes it easier for the field to roll through the steepening feature.

\bibliographystyle{jhep}
\bibliography{bibliography,all}

\end{document}